\documentclass[12pt,a4paper]{article}
\usepackage[DIV=13]{typearea}
\usepackage[onehalfspacing]{setspace}
\usepackage[english]{babel}
\usepackage[ansinew]{inputenc}
\usepackage{amssymb,amsmath,bm,array,dsfont,graphicx,natbib}
\usepackage[colorlinks,citecolor=black,linkcolor=black,urlcolor=black]{hyperref}
\usepackage[hang,tight]{subfigure}
\usepackage{authblk}
\usepackage{framed}
\usepackage{xcolor}
\usepackage{bbm}
\usepackage{longtable}
\usepackage{rotating}
\usepackage{amsthm}

\colorlet{shadecolor}{gray!25}

\newcommand{\newoperator}[3]{\newcommand*{#1}{\mathop{#2}#3}}
\newcommand{\renewoperator}[3]{\renewcommand*{#1}{\mathop{#2}#3}}
\newcommand{\mQ}{Q}

\newcommand{\mR}{R}

\newcommand{\mT}{T}

\newcommand{\vy}{y}
\newcommand{\mZ}{Z}

\newcommand{\valpha}{\alpha}

\newcommand{\veta}{\eta}
\newcommand{\vtheta}{\theta}

\newcommand{\mSigma}{\varSigma}

\renewoperator{\Re}{\mathrm{Re}}{\nolimits}
\renewoperator{\Im}{\mathrm{Im}}{\nolimits}
\makeatletter
\newcommand{\rd}{\@ifnextchar^{\DIfF}{\DIfF^{}}}
\def\DIfF^#1{%
   \mathop{\mathrm{\mathstrut d}}%
   \nolimits^{#1}\gobblespace}
\def\gobblespace{\futurelet\diffarg\opspace}
\def\opspace{%
   \let\DiffSpace\!%
   \ifx\diffarg(%
   \let\DiffSpace\relax
   \else
   \ifx\diffarg[%
   \let\DiffSpace\relax
   \else
   \ifx\diffarg\{%
   \let\DiffSpace\relax
   \fi\fi\fi\DiffSpace}

\newcommand{\E}{\operatorname{E}}

\newcommand{\Var}{\operatorname{Var}}
\newcommand{\Corr}{\operatorname{Corr}}
\newcommand{\Cov}{\operatorname{Cov}}

\newoperator{\ip}{\mathrm{int}}{\nolimits}

\newcommand{\beq}{\begin{equation}}
\newcommand{\eeq}{\end{equation}}
\newcommand{\bal}{\begin{align*}}
\newcommand{\eal}{\end{align*}}

\newcommand{\bvec}{\begin{pmatrix}}
\newcommand{\evec}{\end{pmatrix}}
\newcommand{\bmat}{\begin{bmatrix}}
\newcommand{\emat}{\end{bmatrix}}

\newcommand{\bsmat}{\begin{smallmatrix}}
\newcommand{\esmat}{\end{smallmatrix}}

\usepackage[bottom]{footmisc}

\allowdisplaybreaks
\title{\vspace{-1cm} Fractional trends and cycles in macroeconomic time series}
\author[1,2]{Tobias Hartl\footnote{Corresponding author. E-Mail: tobias1.hartl@ur.de\\
		The authors thank James Morley, the participants of the econometric seminar in Nuremberg, the department seminar at the Christian Albrechts University Kiel, the DAGStat conference 2019 in Munich, and the workshop on high-dimensional time series in economics and finance 2019 in Vienna. Support through the projects TS283/1-1 and WE4847/4-1 financed by the German Research Foundation (DFG) is gratefully acknowledged.}}
\author[1]{Rolf Tschernig}
\author[1,2]{Enzo Weber}
\affil[1]{University of Regensburg, 93053 Regensburg, Germany}
\affil[2]{Institute for Employment Research (IAB), 90478 Nuremberg, Germany}
\date{May 2020}

\begin{document}
\maketitle

\thispagestyle{empty}
\setcounter{page}{0}
\vspace{-1cm}
\paragraph{\bf Abstract.}
\begin{spacing}{1.15}
We develop a generalization of correlated trend-cycle decompositions that avoids prior assumptions about the long-run dynamic characteristics by modelling the permanent component as a fractionally integrated process and incorporating a fractional lag operator into the autoregressive polynomial of the cyclical component. The model allows for an endogenous estimation of the integration order jointly with the other model parameters and, therefore, no prior specification tests with respect to persistence are required. We relate the model to the Beveridge-Nelson decomposition and derive a modified Kalman filter estimator for the fractional components. Identification, consistency, and asymptotic normality of the maximum likelihood estimator are shown. For US macroeconomic data we demonstrate that, unlike $I(1)$ correlated unobserved components models, the new model estimates a smooth trend together with a cycle hitting all NBER recessions. While $I(1)$ unobserved components models yield an upward-biased signal-to-noise ratio whenever the integration order of the data-generating mechanism is greater than one, the fractionally integrated model attributes less variation to the long-run shocks due to the fractional trend specification and a higher variation to the cycle shocks due to the fractional lag operator, leading to more persistent cycles and smooth trend estimates that reflect macroeconomic common sense.
\end{spacing}

\vspace{-0.1cm}

\vspace{-0.1cm}
\paragraph{\bf JEL-Classification.}

C22, C51, E32

\newpage

\section{Introduction}
Unobserved components (UC) models are widely applied in macroeconomic research, e.g.\ to decompose GDP and industrial production into trend and cycle \citep[][]{Har1985, MorNelZi2003, Web2011, MorPig2012}, to study cyclical consumption \citep[][]{Mor2007}, and to measure long-run investment \citep[][]{HarTri2003}. While empirical evidence supports a strong negative correlation of long- and short-run shocks for the aforementioned macroeconomic aggregates, the correlated UC model as proposed by \cite{BalWoh2002} and \cite{MorNelZi2003} frequently produces a volatile long-run component estimate together with a noisy cycle \citep{Web2011, KamMorWo2018}, thereby contradicting macroeconomic common sense. In addition, the integration order of the long-run component is subject to debate. While all aforementioned papers model the long-run component as an $I(1)$ process, \cite{Cla1987} and \cite{OhZiv2006} suggest specifications with $I(2)$ trends that nest the HP filter \citep[cf.][]{Gom1999, Gom2001}, and $I(0)$ specifications with structural breaks are considered in \cite{PerWad2009} and \cite{Wad2012}. 

\noindent
In this paper, we argue that economically implausible cycle estimates are likely to result from a too restrictive specification of the integration order: If a process is in fact integrated of order greater than one, then misspecifying the long-run component to be $I(1)$ upward-biases the variance estimate of the long-run shocks, yielding a high signal-to-noise ratio. This results in a volatile trend estimate together with a noisy cycle. As will be shown, generalizing the persistence properties from integer integration orders to the fractional domain and estimating the integration order jointly with the other parameters of the model can solve the problem.


\noindent

\noindent

\noindent
Focusing on key macroeconomic indicators, the assumption of integer integration orders (0, 1, or 2) has been contested for several variables. 
For real GDP, \cite{MueWat2017} find that the likelihood is flat around $d=1$, yielding a $90\%$ confidence interval for $d$ that is given by $[0.51, 1.44]$. Inference for $d>1$ is found in \cite{Cha1998} for low frequency transformations of income, consumption, investment, exports, and imports for the UK, and in \cite{Erg2019} for the GDP of several high-income OECD countries. Finally, \cite{DieRud1991} find evidence for $d>1$ for consumption and income.

\noindent

\noindent
We contribute to the methodological literature by deriving a fractional trend-cycle decomposition, where the long-run component is allowed to be fractionally integrated ($I(d)$), $d \in \mathbb{R}^+$, whereas the fractional lag operator $L_d = 1-\Delta_+^d$ of \cite{Joh2008}, that is defined in \eqref{diff:op}, enters the lag polynomial of the cyclical component.
In UC models fractionally integrated processes allow for richer dynamics of both, trend and cycle components, and nest a broader class of data-generating mechanisms. Contrary to $I(1)$ UC models, they allow for mean-reversion of the long-run component for $d<1$, while $d>1$ assigns a higher persistence to long-run shocks. 
Particularly in the latter case the variance estimate of the long-run component in $I(1)$ UC models is upward-biased, yielding a high signal-to-noise ratio that causes volatile trend estimates together with noisy, implausible cycles. Conversely, fractionally integrated UC models are likely to adequately capture the true signal-to-noise ratio and produce reliable trend and cycle estimates, as they nest the data-generating mechanism. 

\noindent
In addition, the fractional lag operator $L_d$ yields a weighted sum of past realizations when multiplied to a contemporaneous random variable, and thus it qualifies as a lag operator \citep{Joh2008, TscWebWe2013}. Since it preserves the integration order of a process for non-negative $d$, the fractional lag operator adds flexibility to the short-run properties of a process rather than influencing the integration order. 
While the standard lag operator $L = 1 - (1-L)$ subtracts an $I(-1)$ process from a contemporaneous variable, the fractional lag operator $L_d = 1- (1-L)^d_+$ subtracts an $I(-d)$ process. 
Thus, for equality of the variances of the two generic processes $L z_{1,t} = z_{1, t-1}$, $L_d z_{2,t} = (1-\Delta_+^d)z_{2,t}$, i.e.\ $\Var(z_{1,t-1}) = \Var((1-\Delta_+^d)z_{2,t}) $, with $z_{1,t}, z_{2,t}$ white noise, it must hold that $\Var(z_{1,t}) < \Var(z_{2,t})$ for $d>1$ and vice versa for $d<1$. 
Consequently for $d>1$, which will be the relevant case in our applications, 
the variance of the short-run shocks in the fractionally integrated UC model must be greater than in the $I(1)$ case to arrive at the same variance of the cyclical component, yielding a smaller signal-to-noise ratio and, therefore, a more persistent cycle.

\noindent
Since $d$ is defined on a continuous support and enters the likelihood function as an unknown parameter, our model allows for an endogenous estimation of the integration order jointly with the other model parameters, avoids prior unit root testing and takes into account model selection uncertainty with respect to $d$. 

\noindent
The canonical (reduced) form of the fractional trend-cycle model exhibits a fractional ARIMA representation in the fractional lag operator $L_d$, and directly relates to a generalization of the Beveridge-Nelson decomposition to the fractional domain. We discuss identification and show that consistency and asymptotic normality carries over from the ARFIMA estimator of \cite{HuaRob2011} and \cite{Nie2015}. Contrary to the correlated UC model, that requires an autoregressive cycle of order $p \geq 2$ to uniquely identify trend and cycle, our model is identified for any $p \geq 0$ whenever $d \neq 1$. 
Finally, we assess the state space representation of the fractional trend-cycle model and propose a computationally efficient modification of the Kalman filter for the estimation of the latent long- and short-run components. 

\noindent
When taking the new fractional UC framework to the data, we demonstrate that it makes an important difference for key questions in empirical macroeconomics. We contribute to the empirical literature by applying our fractional trend-cycle decomposition to US GDP, industrial production, private investment, and personal consumption. Our model nests a wide class of UC models and allows to draw inference on the proper specification of the long-run component for the four series under study. We contrast the trend and cycle estimates from integer-integrated UC models with the results from our fractional decomposition and state differences regarding shape, smoothness, variance and importance of the different components. Especially for industrial production we obtain a decomposition that is in line with economic theory, as the cyclical component captures all NBER recession periods, whereas the correlated UC model clearly fails to produce a plausible cycle. For all time series under study, we estimate a continuous increase of the cyclical components in periods of economic upswing, where the correlated UC model produces a noisy cycle that sharply increases before the economy is hit by a recession.

\noindent
The structure of the paper is as follows. Section \ref{Ch:2} details the fractional UC model, relates it to well-established trend-cycle decompositions, derives the reduced form, and discusses identification. Section \ref{sec:ss} derives the state space representation, proposes a computationally efficient modified Kalman filter for the estimation of the latent fractional trend and cycle, and discusses parameter estimation.
In section \ref{sec:app} the model is applied to decompose macroeconomic aggregates. Section \ref{Ch:4} concludes.

\section{A fractional trend-cycle decomposition} \label{Ch:2}
We define a fractional trend-cycle decomposition of a scalar time series $\{y_{t}\}_{t=1}^{n}$ as the sum of a long-run component $\tau_{t}$ and a cycle component $c_{t}$
\begin{align}\label{eq:t}
	y_{t} = {\tau}_{t} + c_{t}, \qquad t = 1,...,n.
\end{align}
The long-run component $\tau_{ t}$ is characterized by an autocovariance function that decays more slowly than with an exponential rate and, therefore, captures the long-run dynamics of a time series, whereas the cycle component $c_{t}$ is $I(0)$ and accounts for transitory fluctuations of a series around its trend. In contrast to the bulk of the literature on unobserved components models that specifies the stochastic trend component typically as a nonstationary process integrated of order 1 or 2 -- most often as a random walk -- we suggest a more general formulation. $\tau_t$ is specified as a combination of a linear deterministic process and a fractionally integrated series
\begin{align}\label{eq:tau}
\tau_{t} = \mu_{0} + \mu_{1} t + x_{t}, \qquad \Delta_+^{d}x_{t} = \eta_{t},
\end{align} 
where $\mu_0$ and $\mu_1$ are constants, $\eta_{t} \sim \mathrm{i.i.d. \ N}(0, \sigma_{\eta}^2)$, and $d \in \mathbb{R}^+$. The fractional difference operator $\Delta^{d}$ is defined as
\begin{align}\label{diff:op}
\Delta^{d} &= (1-L)^{d} = \sum_{j = 0}^{\infty}\pi_{j}(d)L^{j},  \qquad
\pi_{j}(d) = 
\begin{cases}
\frac{j-d-1}{j}\pi_{j-1}(d) &  j = 1, 2, ..., \\ 
1										&   j = 0,
\end{cases} 
\end{align}
and a $+$-subscript denotes a truncation of an operator at $t \leq 0$, e.g. for an arbitrary process $z_t$, $\Delta_+^d z_t= \sum_{j=0}^{t-1}\pi_j(d) z_{t-j}$ \citep[][def.\ 1]{Joh2008}.
The fractional long-run component $x_t$ adds flexibility to the weighting of past shocks for $d \in \mathbb{R}^+$ and nests the classic integer integrated specifications for $d \in \mathbb{N}$. The memory parameter $d$ determines the rate at which the autocovariance function of $x_t$ decays, and a higher $d$ implies a slower decay. For $d<1$ $x_t$ is mean-reverting, while $d \in [1, 2)$ yields the aggregate of a mean-reverting process.
Throughout the paper, we adopt the type II definition of fractional integration \citep{MarRob1999} that assumes zero starting values for all fractional processes, and, as a consequence, allows for a smooth treatment of the asymptotically stationary ($d < 0.5$) and the nonstationary ($d \geq 0.5$) case. Due to the type II definition of fractional integration the inverse of the fractional difference operator $\Delta_+^{-d} = (1-L)^{-d}_+$ exists for all $d$, such that we can write
\begin{align*}
	x_{t} = \Delta_+^{-d}\eta_t=\sum_{j=0}^{t-1} \varphi_j(d) \eta_{t-j} \qquad
	\varphi_{j}(d) = 
	\begin{cases}
	\frac{j+d-1}{j}\varphi_{j-1}(d) &  j = 1, 2, ..., \\ 
	1										&   j = 0.
	\end{cases} 
\end{align*}

\noindent
Turning to the transitory component, we allow for an AR($p$) process in the fractional lag operator
 \begin{align}\label{eq:c}
 	\phi(L_{d})c_{t}=\varepsilon_{t},
 \end{align}
 where $\phi(L_{d})=1-\phi_{1}L_{d} - ... - \phi_{p} L_{d}^p$, $L_{d} = 1-\Delta_+^{d}$ is the fractional lag operator \citep[][eq.\ 2]{Joh2008}, and $\varepsilon_{t} \sim \mathrm{i.i.d. \ N}(0, \sigma_{\varepsilon}^2)$. For stability of the fractional lag polynomial $\phi(L_{d})$ the condition of \citet[cor.\ 6]{Joh2008} is required to hold. It implies that the roots of $\vert \phi(z) \vert = 0$ lie outside the image $\mathbb{C}_{d}$ of the unit disk under the mapping $z \mapsto 1-(1-z)^{d}$. In fractional models $L_d$ plays the role of the standard lag operator $L_1 = L$, since $(1-L_d) x_t= \Delta_+^d x_t \sim I(0)$. For an arbitrary process $z_t$, $L_d z_t = - \sum_{j=1}^{t-1}\pi_j(d)z_{t-j}$ is a weighted sum of past $z_t$, and hence $L_d$ qualifies as a lag operator.  
 Furthermore, by definition the filter $\phi(L_{d})$ preserves the integration order of a series since $d > 0$.
 
 \noindent
 We do not exclude contemporaneous correlation between trend and cycle innovations. Hence, we allow $\rho = \Corr(\eta_t, \varepsilon_t) \neq 0$, which directly implies $\mathrm{E}(\eta_{t} \varepsilon_{t}) = \sigma_{\eta \varepsilon}\neq 0$. For different time indexes we restrict the cross-correlation to be zero, $\mathrm{E}(\eta_{t} \varepsilon_{s}) = 0 \ \forall t \neq s$. Thus, the long- and short-run shocks are i.i.d.\ N distributed with non-diagonal variance $Q$
 \begin{align*}
 	\begin{pmatrix}
		\eta_{t} \\ \varepsilon_{t} 
 	\end{pmatrix} \sim  \mathrm{i.i.d.\ N}(0, \mQ), \qquad 
	 \mQ = \begin{bmatrix}
	 	\sigma_\eta^2 & \sigma_{\eta \varepsilon} \\
	 	\sigma_{\eta \varepsilon} & \sigma_\varepsilon^2
 \end{bmatrix}.
 \end{align*}

 \noindent
 Our model is very general in terms of its long-run dynamic characteristics, as it nests the well-known framework of \cite{Har1985} for $d=1$, where the long-run component is a random walk with drift, and $c_{t}$ is an autoregressive process of finite order. Correlated shocks as in \cite{BalWoh2002}, \cite{MorNelZi2003}, and \cite{Web2011} are explicitly allowed. For $d=2$, one obtains the double-drift unobserved components model of \cite{Cla1987}, and a fractional plus noise decomposition as proposed in \cite{Har2002} is obtained by setting $d \in \mathbb{R}^+$, $p=0$. 

 \noindent
As shown in \eqref{eq:farima} below, similar to the classic UC-ARMA model that exhibits an ARIMA representation \citep[][eq.\ 2b]{MorNelZi2003} the canonical form of our fractional trend-cycle model is an ARIMA($p, d, n-1$) model in the fractional lag operator. To see this, note that $(1- L_{d})[\phi(L_{d})^{-1}]_+ \varepsilon_{t} = \theta^\varepsilon_+(L_d) \varepsilon_t$ is a stable moving average process in the fractional lag operator \citep[cf.][]{Joh2008}, such that we can write
 \begin{align}
 	\Delta^{d}_+(y_{t} - \mu_{0}-\mu_{1} t )&=  \eta_{t} + (1- L_{d})[\phi(L_{d})^{-1}]_+ \varepsilon_{t}= \theta_+^u(L_{d})u_{t}, \label{eq:fBN} \\
\phi(L_{d})\Delta^{d}_+(y_{t} - \mu_{0}-\mu_{1} t )&= \phi(L_{d} )\eta_{t} + (1- L_{d})\varepsilon_{t}= \psi_+(L_{d})u_{ t} \label{eq:farima}, 
\end{align} 
where $\psi_+(L_d) = \phi(L_d)\vtheta_+^u(L_d)$ is a truncated moving average polynomial of infinite order that results from the aggregation of $\phi(L_{d} )\eta_{t} + (1- L_{d})\varepsilon_{t}$. Its existence together with a recursive formula for the coefficients $\psi_j$ is shown in appendix \ref{App:MA_agg}. 
$u_{t} \sim \mathrm{N}(0, \sigma_u^2)$ holds the disturbances and is Gaussian white noise with $\sigma_u^2 = \sigma_\eta^2 + \sigma_\varepsilon^2 + 2\sigma_{\eta \varepsilon}$, which follows from \citet[][p.\ 248f]{GraMor1976} for contemporaneously dependent $\varepsilon_t$, $\eta_t$, and $\theta_+^u(L_d) = \sum_{i=0}^{t-1}\theta_i^u L_d^i$, $\vtheta_0 = 1$, $\theta_i^u = \frac{\sigma_\varepsilon}{\sigma_u}\theta_i^\varepsilon $ for all $i > 0$. While aggregating MA processes in the standard lag operator yields an MA process whose lag length equals the maximum lag order of its aggregates, this does not hold in general for the aggregation of MA processes in the fractional lag operator $L_d$, since $L_d^i u_t$, $L_d^j u_t$ are not independent for $i, j > 1$, $i \neq j$. Only for $p=1$ equation \eqref{eq:farima} becomes an ARIMA($1, d, 1$) in the fractional lag operator, since $\eta_t + \varepsilon_t$ and $L_d (\phi_1 \eta_t + \varepsilon_t)$ are independent. For $d \in \mathbb{N}$ the model in \eqref{eq:farima} nests the integer-integrated ARIMA models. Due to the inclusion of the fractional lag operator \eqref{eq:farima} differs from the standard ARFIMA model. Nonetheless, \eqref{eq:farima} exhibits an ARFIMA($n-1$, $d$, $n-1$) representation as $\phi(L_d)$ can be written as an AR($n-1$) polynomial.



\noindent
Our fractional trend-cycle model can be seen as a generalization of the decomposition of \cite{BevNel1981} to the fractional domain. To see this, consider \eqref{eq:fBN} from which one obtains directly
\begin{align*}
	(1-L_{d}) (y_{t} - \mu_{0} - \mu_{1}t) &=  \vtheta_+^u(L_{d}) u_{t} = \theta_+^u(1) u_{t} - (1-L_{d}) \sum_{k=0}^{t-2} L_{d}^k u_{t} \sum_{j=k+1}^{t-1}\theta_{j}^u,
\end{align*} 
such that multiplication with $\Delta_+^{-b}$ yields the long- and short-run components 
\begin{align}\label{eq:BN}
	x^{BN}_{t} = \Delta^{-d}_+ \vtheta_+^u(1) u_{t} = x_t, \qquad c_{t}^{BN} = - \sum_{k=0}^{t-2} L_{d}^k u_{t} \sum_{j=k+1}^{t-1}\theta_{j}^u =c_t.
\end{align}
 Equality of the decomposition of \cite{BevNel1981} and the UC model in \eqref{eq:t}, \eqref{eq:tau}, and \eqref{eq:c} was shown in \cite{MorNelZi2003} for $d=1$. Note that $x_t^{BN}$ and $c_t^{BN}$ are identical to the unobserved components in \eqref{eq:tau} and \eqref{eq:c} for any $d$, which follows immediately from plugging $L_d = 1$ in \eqref{eq:fBN}. Consequently, the fractional trend-cycle decomposition generalizes the $I(1)$ Beveridge-Nelson decomposition to the class of ARIMA models in the fractional lag operator. 

\noindent
For $d = 1$ \cite{MorNelZi2003} demonstrate that the integer-integrated UC model is not identified for $p=1$, $d= 1$, $\sigma_{\eta \varepsilon}\neq 0$. In that case, imposing the restriction $\sigma_{\eta \varepsilon}=0$ yields a  decomposition that is different to the one of \cite{BevNel1981}. The same is shown by \cite{Web2011} for the simultaneous unobserved components model identified by heteroscedasticity and by \cite{TreWeb2016} for the multivariate UC model.

\noindent
In fact, $d= 1$ is the only case where the unobserved components model is not identified for $p = 1$. In any other case, where $d \in \mathbb{R}^+$, $d\neq 1$, we show in the following that the model parameters $\phi(L_{d})$, $d$, $\sigma_{\eta}$, $\sigma_{\varepsilon}$, and $\sigma_{\eta \varepsilon}$ can be uniquely recovered from \eqref{eq:farima}, which is sufficient for identification. Since $\phi(L_{d})$ and $d$ are obtained directly from the model in its canonical form,
we consider $\sigma_{\eta}$, $\sigma_{\varepsilon}$, and $\sigma_{\eta \varepsilon}$ on which identification of the unobserved components model crucially depends.

\noindent

\noindent
For $d \neq 1$ the parameters $\sigma_{\eta}$, $\sigma_{\varepsilon}$, and $\sigma_{\eta \varepsilon}$ are obtained from the autocovariance function of $\psi(L_d)u_{t}$ for any $p\geq 0$, whereas $p \geq 2$ is required for $ d=1$, as \cite{MorNelZi2003} demonstrate. To see this, we consider $p=2$, for which
\begin{align*}
	\gamma_0 = \Var \left[\psi(L_{d})u_{t}\right] &= \sigma_\eta^2 \left\{ 1 + \sum_{k=1}^{t-1}\left[ (\phi_{1} + 2 \phi_{2})\pi_k(d) - \phi_{2} \pi_k(2d) \right]^2 \right\}  + \sigma_\varepsilon^2 \sum_{k=0}^{t-1} \pi_k(d)^2 \\
	&+2 \sigma_{\eta \varepsilon} \left\{ 1 +  \sum_{k=1}^{t-1}\left[ (\phi_{1} + 2 \phi_{2})\pi_k(d) - \phi_{2} \pi_k(2d) \right]\pi_k(d) \right\}, 
	\end{align*}
\begin{align*}
	\gamma_j &=\Cov [ \psi(L_{d})u_{t}, \psi(L_{d})u_{t-j}]= \sigma_\eta^2 \Big\{ \left[(\phi_{1} + 2\phi_{2}) \pi_j(d) - \phi_{2} \pi_j(2d)\right] \\
	&+ \sum_{k=j+1}^{t-1} \left[  (\phi_{1} + 2\phi_{2})\pi_k(d) - \phi_{2} \pi_{k}(2d) 
	\right] \left[  (\phi_{1} + 2\phi_{2})\pi_{k-j}(d) - \phi_{2} \pi_{k-j}(2d) 
	\right]
	\Big\}
	\\
	&+ \sigma_\varepsilon^2 \sum_{k=j}^{t-1} \pi_k(d) \pi_{k-j}(d) + \sigma_{\eta \varepsilon} \Bigg\{ \pi_j(d) +  \sum_{k=j}^{t-1}\left[ (\phi_{1} + 2 \phi_{2})\pi_k(d) - \phi_{2} \pi_k(2d) \right]\pi_{k-j}(d) \\
	&+ \sum_{k=j+1}^{t-1} \left[ (\phi_{1} + 2 \phi_{2})\pi_{k-j}(d) - \phi_{2} \pi_{k-j}(2d) \right]\pi_{k}(d) 
	\Bigg\}.
\end{align*} 
Note that $d = 1$ implies $\pi_j(d)=0 \ \forall j>1$ and $\pi_j(2d)=0 \ \forall j > 2$. If now $\phi_{2} = 0$, then $\gamma_j=0 \ \forall j >1$, and hence the model is not identified, as also \cite{MorNelZi2003} demonstrate. For $d \neq 1$ the model is identified for any $p\geq0$, since $\gamma_{2} \neq 0$. 
Contrary to the $I(2)$-model of \cite{OhZivCr2008} with three shocks that requires $p \geq 4$,
our model is identified for any $p \geq 0$ when $d=2$, since $\gamma_0$, $\gamma_1$, $\gamma_2$ are different from $0$ which is sufficient for the identification of $\sigma_{\eta}$, $\sigma_{\varepsilon}$, and $\sigma_{\eta \varepsilon}$.

 
\section{State space representation and estimation}\label{sec:ss}
In this section we derive a state space representation for the fractional trend-cycle decomposition together with a modified Kalman filter estimator for the unobserved components. Furthermore, we discuss the maximum likelihood estimator for the unknown model parameters and show that consistency carries over from the ARFIMA estimator of \cite{HuaRob2011} and \cite{Nie2015}.

\noindent
There exists a finite-order state space representation of \eqref{eq:tau} and \eqref{eq:c} for fixed sample size $n$, since any type II fractionally integrated process exhibits an autoregressive representation of order $n-1$. 
Thus, an exact state space representation of the fractionally integrated system yields a state vector of dimension $k \geq 2n$ for $d \notin \mathbb{N}$. Estimation of $\tau_t$, $c_t$ via the Kalman filter then involves the inversion of the $k \times k$ conditional state variance for each $t=1,...,n$, which slows down the Kalman filter substantially for large $n$.  

\noindent
Consequently, different approximations for fractionally integrated processes in state space form have been considered \citep[cf.\ e.g.][for truncated MA and AR approximations]{ChaPal1998, Pal2007}. \cite{HarWei2018} study ARMA($v$, $w$) approximations with $v, w \in \{2, 3, 4\}$ for fractional processes. \cite{HarTscWeb2019} then correct for the resulting approximation error of the ARMA approximations for fractionally integrated trends. Their approach keeps the state dimension manageable and is feasible from a computational perspective, as it only requires to calculate the inverse of $\Var[(y_{1}, ...., y_n)']$ once. Furthermore, it yields the identical likelihood function as the exact state space model but is computationally superior.
We generalize their method to the fractional trend-cycle model in the following, and thereby provide a computationally feasible exact Kalman filter estimator for $\tau_t$, $c_t$. 

\noindent
To begin with,  collect all model parameters in $\vtheta = (d, \phi_1, ..., \phi_p, \sigma_\eta, \sigma_{\eta \varepsilon}, \sigma_{\varepsilon})'$ and define $\E_\vtheta$, $\Var_\vtheta$, $\Cov_\vtheta$ as the moments given the parameter vector $\vtheta$. Define $\mathcal{F}_t$ as the $\sigma$-field generated by $y_1, ...., y_t$ and let $z_{t|s} = \E_\vtheta(z_t| \mathcal{F}_s )$ for any random variable $z_t$. Then the prediction error of the Kalman filter for the exact state space model of \eqref{eq:t}, \eqref{eq:tau}, and \eqref{eq:c} is
\begin{align}\label{eq:vt}
	v_{t+1} = y_{t+1} - \E_\vtheta (\vy_{t+1} | \mathcal{F}_t) = y_{t+1} - \mu_0 - \mu_1 (t+1) - x_{t+1|t} - c_{t+1|t}.
\end{align}
Let $\tilde{x}_{t}$ and $\tilde{c}_t$ denote the approximate long-run and cyclical components defined in detail in \eqref{approximation:x} and \eqref{approximation:c} below. We will show that the following relationships for the conditional expectations hold
\begin{align}
	\tilde{x}_{t+1|t} &= x_{t+1|t} - \epsilon_{t}^x, \label{apx}\\
	\tilde{c}_{t+1|t} &= c_{t+1|t} - \epsilon_{t}^c, \label{apc}
\end{align}
where $\epsilon_{t}^x$, $\epsilon_{t}^c$ denote the approximation errors of the Kalman filter estimates and are $\mathcal{F}_t$-measurable. Then \eqref{eq:vt} can be rewritten as
\begin{align}\label{eq:v}
v_{t+1}=  y_{t+1} - \mu_0 - \mu_1(t+1) - \epsilon_{t}^x - \epsilon_{t}^c - \tilde{x}_{t+1|t} - \tilde{c}_{t+1|t} = \ddot{y}_{t+1} - \E_\vtheta (\ddot{y}_{t+1}|\mathcal{F}_t),
\end{align}
with $\ddot{y}_{t+1}= y_{t+1} - \epsilon_{t}^x - \epsilon_{t}^c$ as the approximation-corrected $y_t$. Therefore, the prediction errors of the approximation-corrected model and the exact state space model are the same. Next we derive \eqref{apx} and \eqref{apc}.

\noindent
The long-run component $x_t$ in \eqref{eq:tau} is approximated by an ARMA($v, w$) process $\tilde{x}_{t} = [a(L, d)^{-1}m(L, d)]_+\veta_t = \sum_{j=0}^{t-1} b_{j}(d)\veta_{t-j}$ where $a_0=m_0=1$ and, therefore, $b_0 = 1$. $a(L, d)$ is an AR polynomial of order $v$, whereas $m(L, d)$ is a MA polynomial of order $w$. This yields an approximation error  
\begin{align} \label{approximation:x}
	x_{t} - \tilde{x}_{t} = \sum_{j=1}^{t-1}(\varphi_{j}(d) - b_{j}(d))\veta_{t-j}. 
\end{align}
The ARMA coefficients are obtained beforehand by minimizing the mean squared error between the Wold representations $x_{t} = \sum_{j=0}^{t-1}\varphi_{j}(d)\veta_{t-j}$ and $\tilde{x}_{t} = \sum_{j=0}^{t-1} b_j(d)\veta_{t-j}$ for a fixed $d$ and sample size $n$. A continuous function that maps from the integration order $d$ to its ARMA coefficients is then obtained by optimizing over a grid of $d$ and smoothing the outcomes using splines. Hence, optimization of the likelihood for the fractional trend-cycle decomposition is conducted over the scalar fractional integration order $d$, and does not involve the estimation of any parameters in $a(L, d)$, $m(L, d)$, such that the dimension of the parameter vector $\vtheta$ is kept small during the optimization. 
Details together with a large simulation study are contained in \cite{HarWei2018}. 

\noindent
The cyclical component $c_{t}$ can be expressed as an AR($n-1$) process in the standard lag operator $\phi(L_{d})c_{t} = \delta_+(L, d, \phi)c_{t}$ that is initialized deterministically with $c_{t}=0$ $\forall t \leq 0$ and where $\delta_+(L, d, \phi)$ results from
\begin{align*}
	\phi(L_{d})c_{t} &=  \left[\sum_{j=0}^{p}\phi_{j}\left( -\sum_{k=1}^{\infty} \pi_{k} (d)L^k  \right)^j  \right]_+ c_{t} = \sum_{j=0}^{t-1} \delta_{ j} c_{ t-j}.
\end{align*}
An approximation for the fractional cyclical component is obtained by truncating $\delta(L, d, \phi)$ after lag $l$, $\tilde{\delta}(L, d, \phi)=\sum_{j=0}^{l} {\delta}_{j}L^j$, $\tilde{\delta}(L, d, \phi) \tilde{c}_{t} = \varepsilon_{t}$. Note that $\delta(L, d, \phi)$, $\tilde{\delta}(L, d, \phi)$ solely depend on $d$ and $\phi_1,...,\phi_p$.
Define $\omega_+(L, d, \phi)=[\delta(L, d, \phi)]^{-1}_+$, $\tilde{\omega}_+(L, d, \phi)=[\tilde{\delta}(L, d, \phi)]^{-1}_+$ as moving average lag polynomials of $c_{t}$, $\tilde{c}_{t}$ in the standard lag operator $L$. The approximation error is then given by
\begin{align}\label{approximation:c}
c_{t}  - \tilde{c}_{t} &= \left[  \left(\sum_{j=0}^{t-1}\delta_{j}L^j\right)^{-1}_+  - \left(\sum_{j=0}^{l}\delta_{j}L^j\right)^{-1}_+ \right] \varepsilon_{t} = \sum_{j=0}^{t-1} \left(  \omega_{j} - \tilde{\omega}_{j} \right) \varepsilon_{ t-j}.
\end{align}

\noindent
With these approximations and an expression for the resulting approximation error at hand, we are ready to derive the impact of the two approximations on the Kalman filter estimates of $\tau_{t}$ and $c_{t}$. 
Note that 
\begin{align}
	\Cov_\vtheta (y_{t}, \eta_{t-j}) &= \varphi_j(d)\sigma_\eta^2 + \omega_{j} \sigma_{\eta \varepsilon} ,  \label{Cov:1}\\
	\Cov_\vtheta (y_{t}, \varepsilon_{t-j}) &= \varphi_{j}(d) \sigma_{\eta \varepsilon} + \omega_{j} \sigma_{\varepsilon}^2,\label{Cov:2}\\
	\Cov_\vtheta (y_{t}, y_{t-j}) &= \sum_{k=0}^{t-j-1}  \varphi_k(d)\varphi_{k+j}(d) \sigma_\eta^2 + \sum_{k=0}^{t-j-1}(\omega_{k} \varphi_{k+j}(d) + \varphi_k(d) \omega_{k+j}) \sigma_{\eta \varepsilon}  \nonumber \\
	&+\sum_{k=0}^{t-j-1}\omega_{k}\omega_{k+j} \sigma_\varepsilon^2, \label{Cov:3}
\end{align}
and define $\vy_{1:t}=(y_{1}, ..., y_{t})' - (\E_\vtheta[y_{1}], ..., \E_\vtheta[y_{t}])' $, $\eta_{1:t} = (\eta_{1},...,\eta_{t})'$, and $\varepsilon_{1:t}=(\varepsilon_{1}, ... , \varepsilon_{t})'$. Then the joint distribution can be stated as
\begin{align}
	\begin{pmatrix}
		\eta_{1:t} \\ \varepsilon_{1:t} \\ \vy_{1:t} 
	\end{pmatrix} \sim N \left(0, \begin{bmatrix}
	\sigma_\eta^2 I & \sigma_{\eta \varepsilon} I & \mSigma_{\eta_{1:t} y_{1:t}}\\
	\sigma_{\eta \varepsilon} I & \sigma_{\varepsilon}^2 I &  \mSigma_{\varepsilon_{1:t} y_{1:t}} \\
	 \mSigma_{\eta_{1:t} y_{1:t}}' &  \mSigma_{\varepsilon_{1:t} y_{1:t}}' & \Sigma_{y_{1:t}}
	\end{bmatrix}\right),
\end{align}
where $\Sigma_{\eta_{1:t}y_{1:t}}= \Cov_\vtheta(\eta_{1:t}, y_{1:t})$, $\Sigma_{\varepsilon_{1:t}y_{1:t}}= \Cov_\vtheta(\varepsilon_{1:t}, y_{1:t})$, and $\Sigma_{y_{1:t}}= \Var_\vtheta(y_{1:t})$ with entries from equations \eqref{Cov:1}, \eqref{Cov:2}, and \eqref{Cov:3}. \\
Let $e_{j}$ be a $t$-dimensional unit vector with a one at column j and zeros elsewhere. Then computing the conditional expectation for \eqref{approximation:x} and \eqref{approximation:c} respectively delivers
\begin{align*}
	\tilde{x}_{t+1 | t} &= x_{t+1 | t} + \E_\vtheta [\tilde{x}_{t+1} - {x}_{t+1} | \mathcal{F}_t] = x_{t+1|t } - \sum_{j=1}^{t}(\varphi_{j}(d) - b_{j}(d))\E_\vtheta [ \veta_{t+1-j} | \mathcal{F}_t ] = \\
	&=  x_{t+1 | t}  - \sum_{j=1}^{t}(\varphi_{j}(d) - b_{j}(d)) e_{ t+1-j} \Sigma_{\eta_{1:t}y_{1:t}} \Sigma_{y_{1:t}}^{-1} y_{1:t} = x_{t+1|t} - \epsilon_{t}^x, \\
	\tilde{c}_{t+1|t} &= c_{t+1|t} +  \E_\vtheta [\tilde{c}_{t+1} - {c}_{t+1} | \mathcal{F}_t] = c_{t+1|t} - \sum_{j=1}^{t}(\omega_{j} - \tilde{\omega}_{j}) \E_\vtheta (\varepsilon_{t+1-j} | \mathcal{F}_t) = \\
	&= c_{t+1|t} - \sum_{j=1}^{t}(\omega_{j} - \tilde{\omega}_{j}) e_{t+1-j} \Sigma_{\varepsilon_{1:t}y_{1:t}} \Sigma_{y_{1:t}}^{-1} y_{1:t}=c_{t+1|t} - \epsilon_{ t}^c,
\end{align*}
where $\epsilon_{t}^x$, $\epsilon_{t}^c$ are the approximation errors in \eqref{apx}, \eqref{apc}, and the last step follows from lemma 1 in \cite{DurKoo2012}. It is easy to see that the approximation errors $\epsilon_t^x=\sum_{j=1}^{t}(\varphi_{j}(d) - b_{j}(d)) e_{ t+1-j} \Sigma_{\eta_{1:t}y_{1:t}} \Sigma_{y_{1:t}}^{-1} y_{1:t}$ and $\epsilon_t^c = \sum_{j=1}^{t}(\omega_{j} - \tilde{\omega}_{j}) e_{t+1-j} \Sigma_{\varepsilon_{1:t}y_{1:t}} \Sigma_{y_{1:t}}^{-1} y_{1:t}$ solely depend on the parameters $\vtheta$ and $y_1,...,y_t$. Hence, they are $\mathcal{F}_t$-measurable and can be computed precisely.

\noindent
Define the approximation-corrected $\ddot{y}_{t+1}= y_{t+1} - \epsilon_{t}^x - \epsilon_{t}^c$. Then the prediction error of the exact state space model and of the approximation-corrected, truncated state space model are identical, which proves \eqref{eq:v}. Thus they have the same conditional log likelihood given a set of parameters $\vtheta$. Consequently, maximization of the conditional log likelihood of the approximation-corrected truncated model solves the same optimization problem as for the exact state space representation but reduces the dimension of the state vector. 
The state space representation of the approximation-corrected truncated model is derived in appendix \ref{App:sta}.

\noindent
The exact choice of $v$, $w$ for the ARMA approximation of the fractional trend component and $l$ for the truncation of the fractional lag operator does not affect the equality in \eqref{eq:v}, since the approximation-correction yields the exact likelihood function that is identical with a non-truncated model but is computationally superior. Nonetheless, numerical optimization takes longer when $v$, $w$, $l$ are chosen too big, as the Kalman filter then has to invert high-dimensional covariance matrices. As a rule-of-thumb, we suggest $v=w=4$, which keeps the dimension of the state vector small and is found to resemble the dynamics of fractionally integrated Gaussian noise $\Delta_+^d \eta_t$ well, as it yields a better fit than autoregressive and moving average processes of order $50$  \citep[][]{HarWei2018}. For the cyclical component we suggest $l=10$. We use this specification in all empirical applications that follow.

\noindent
Finally we comment on the estimation of $\vtheta$ via maximum likelihood (ML). Under the prerequisites derived in \citet[Assumptions A.1-A.3]{HuaRob2011} the conditional sum-of-squares estimator of \eqref{eq:farima}, that is asymptotically equivalent to the ML estimator, is consistent and asymptotically normally distributed. As they show, imposing stationarity and invertibility on $\theta_+^u(L_{d})$ together with $u_t$ being white noise is sufficient for consistency and asymptotic normality. Similar results are obtained by  \cite{Nie2015}. Since the ML estimator has the same limit distribution as the conditional sum-of-squares estimator, and since our model in its reduced form satisfies the conditions of \cite{HuaRob2011} and \cite{Nie2015}, their asymptotic results hold for the ML estimator of the reduced form in \eqref{eq:farima}.

\noindent
Under identification, the asymptotic results carry over from the reduced form in \eqref{eq:farima} to the structural form in \eqref{eq:t}, \eqref{eq:tau}, and \eqref{eq:c}. As shown in section \ref{Ch:2}, the parameters of the structural form $\vtheta$ can be uniquely recovered from \eqref{eq:farima} for any $p$ if $d \neq1$ (and for any $p \geq 2$ if $d=1$ as shown in \cite{MorNelZi2003}). Therefore, the maximum likelihood estimator of $\vtheta$ based on the probability density function of the prediction errors in \eqref{eq:vt} is consistent.

\noindent
Consequently, three different model formulations of the fractional trend-cycle decomposition yield consistent estimates for $\vtheta$ in \eqref{eq:t}, \eqref{eq:tau}, and \eqref{eq:c} via maximum likelihood, namely the reduced form model in \eqref{eq:farima}, the exact state space representation based on $\vy_t$ and the approximation-corrected truncated model based on $\ddot{y}_t$. The latter model allows to estimate $\tau_t$ and $c_t$ directly via the Kalman filter and is computationally superior to the exact state space representation. Therefore, it forms the basis of our empirical analysis in the next section. 
\section{Empirical applications} \label{sec:app}
We apply our fractional trend-cycle decomposition in \eqref{eq:t}, \eqref{eq:tau}, and \eqref{eq:c} to extract long-run and transitory components from real GDP, industrial production, gross private domestic investment, and personal consumption expenditures for the US. 
Trend-cycle decompositions of real economic output are typically conducted to estimate the cyclical deviation of output from its long-run growth path. Examples are \citet[][]{HarTri2003, GarRobWr2006, PerWad2009} for log US real GDP and \citet[][]{Cla1987, StoWat1999, Web2011} for log US real industrial production. \citet[][]{Mor2007} estimates the long-run component of personal consumption, whereas \citet{HarTri2003} also consider US investment. Hence, our results from the fractional trend-cycle model can easily be compared and checked against widely used alternatives.

\noindent
In our application several advantages of the fractional trend-cycle decomposition become apparent. 
From a methodological perspective the endogenous treatment of the integration order 
neither requires assumptions about the persistence of a series nor prior unit root testing or differencing. Furthermore, fractional trends offer additional flexibility in modelling the permanent component, which directly affects the estimation of the transitory cycle. 
From an empirical perspective, we contribute to the literature by providing new insights on the persistence of long-run output, investment and consumption, when the trend component is not restricted to be $I(1)$. In addition, we study cyclical adjustments during economic recessions and comment on the correlation structure between permanent and transitory shocks. Finally, we investigate how establishing the fractional lag operator affects the estimate of the cyclical component. Since $d > 0$, $L_d \varepsilon_t$ is a weighted sum of past $\varepsilon_t$ that is $I(0)$. Consequently, the fractional lag operator allows for a more flexible way of modelling the short-run properties of a series while preserving the integration order. 

\noindent
The data was downloaded from the Federal Reserve Bank of St.\ Louis (mnemonics: GDPC1, INDPRO, PCECC96, GPDIC1), is in quarterly frequency and spans from 1961:1 to 2018:4. All series are seasonally and inflation adjusted and enter the dataset in logs.

\noindent
\noindent
To estimate the unknown parameters $\vtheta$ in \eqref{eq:t}, \eqref{eq:tau}, and \eqref{eq:c} we draw $100$ combinations of  starting values from uniform distributions with appropriate support and maximize the log likelihood of the fractional trend-cycle model via the Nelder-Mead algorithm up to a certain relative tolerance.
We ignore the approximation-correction, that has a negligible impact on the performance of the ML estimator as shown in \cite{HarWei2018}, for the estimation of the starting values to speed up the computations.
Next, the parameters corresponding to the greatest log likelihood are set as starting values for a finer maximization via the exact approximation-corrected method discussed in section \ref{Ch:2}. $p$ is chosen via the Bayesian Information Criterion (BIC).

\noindent
To study the impact of fractional trends and cycles we introduce a benchmark model that restricts $d = 1$ in \eqref{eq:tau} and \eqref{eq:c}. Hence, we contrast the fractional trend-cycle model with the $I(1)$ correlated unobserved components model studied in \cite{MorNelZi2003} and \cite{Web2011}. The restricted model is given in equation \eqref{eq:res} below and will be called T-C specification in the following. We will refer to the unrestricted model, that is given in \eqref{eq:unres}, as FT-FC specification 
\begin{align}
	\tau^{(T)}_{t} &= \mu_{0}^{(T)} + \mu_{1}^{(T)}t + \Delta_+^{-1} \eta_{t}^{(T)}, && \phi^{(C)}(L)c^{(C)}_{t} = \varepsilon_{t}^{(C)},\label{eq:res} \\
	\tau^{(FT)}_{t} &= \mu_{0}^{(FT)} + \mu_{1}^{(FT)}t + \Delta_+^{-d} \eta_{t}^{(FT)}, && \phi^{(FC)}(L_{d})c^{(FC)}_{ t} = \varepsilon_{ t}^{(FC)}. \label{eq:unres}
\end{align}
 Both models allow for correlated permanent and transitory shocks, $\rho= \mathrm{Corr}(\eta_{ t}, \varepsilon_{ t})\neq 0$. 

\noindent
Estimation results together with the log likelihoods are reported in table \ref{ta:uni}. 

\subsection*{GDP: gradual cyclical upswing}
\noindent
For log GDP, empirical evidence for the exact value of the persistence parameter $d$ is mixed. \citet[][]{DieRud1989} and \citet{TscWebWe2013} estimate $d$ to be slightly smaller than one, whereas \cite{MueWat2017} find that the likelihood is flat around $d = 1$, such that a $90\%$ confidence interval yields $d \in [0.51, 1.44]$. From the exact local Whittle estimator of \cite{ShiPhi2005} and the method of \cite{GewPor1983} we obtain $\hat{d}^{EW} = 1.24$ and $\hat{d}^{GPH} = 1.24$ with tuning parameter $\alpha = 0.65$ as in \cite{ShiPhi2005}. 

\noindent
For the fractional trend-cycle model the ML estimator yields $\hat{d}^{FT-FC} = 1.32$, implying that log US real GDP is a non-stable, nonstationary fractional process. As figure \ref{fig:ll_gdp} shows, the log likelihood is considerably flat around $\hat{d}^{FT-FC}$, which explains the different results for the persistence parameter in the literature and confirms the findings in \cite{MueWat2017}. Nonetheless, most of the probability mass clearly lies at $d \geq 1$. Contrary to the benchmark, the FT-FC specification attributes more volatility to the transitory shocks, whereas $\sigma_\eta$ is estimated to be smaller than in the T-C specification. 

\begin{figure}[h]
	\includegraphics[scale=0.82, trim = {0cm, 1.2cm, 0cm, 0cm}]{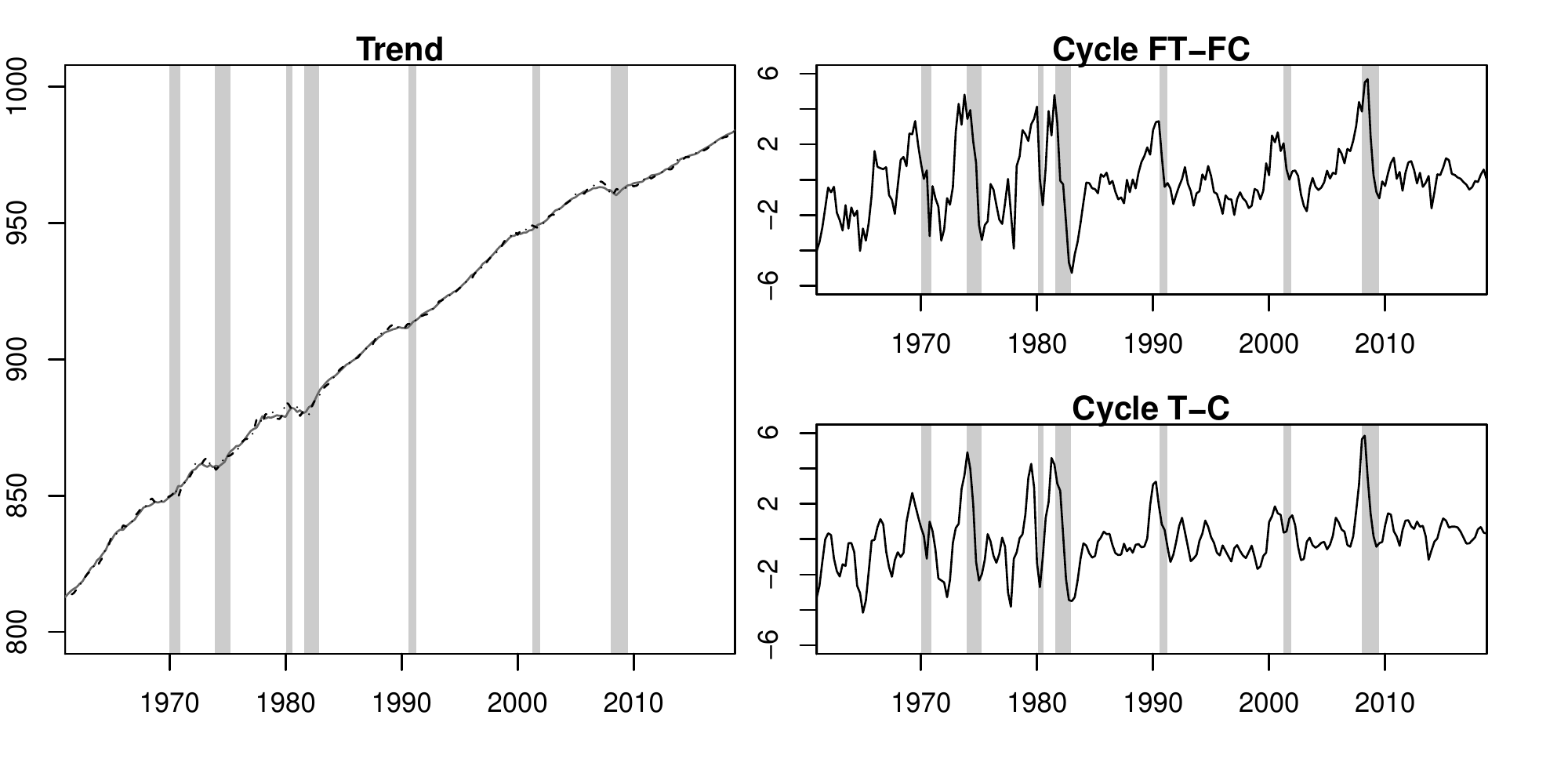}
	\caption[GDPC1: Trend Cycle]{Trend-cycle decompositions for log US real GDP with correlated innovations. The left plot sketches the trend component estimate from the restricted model \eqref{eq:res} (T-C, with $d=1$) in black, dashed, together with the trend component from the unrestricted model \eqref{eq:unres} (FT-FC, with $d \neq 1$ allowed) in gray, solid. The plots on the right-hand side show the cyclical components for the unrestricted and the restricted model. Shaded areas correspond to NBER recession periods.}
	\label{fig:gdp}
\end{figure}%

\noindent
Figure \ref{fig:gdp} plots the decompositions from the T-C and the FT-FC specification in \eqref{eq:res} and \eqref{eq:unres}. At first glance, it demonstrates that the T-C and the FT-FC decomposition for log US real GDP yield rather similar results, which may be due to the flat likelihood of the FT-FC model around $d=1.32$, and the results coincide with the literature \citep[cf.\ e.g.][]{MorNelZi2003, Sin2009}.
As economic theory suggests, both cyclical components decline during the NBER recession periods. The FT-FC specification suggests a gradual cyclical upswing in non-recession periods and therefore captures an important feature of the business cycle, contrary to the cyclical T-C component that exhibits a steep increase right before a recession period. Similar cycle estimates as from the FT-FC specification are obtained from the nonlinear regime-switching UC-FP-UR model of \cite{MorPig2012}. Thereby, the parsimonious parametrization of the FT-FC model together with its ability to resemble nonlinear dynamics foster its generality.
Furthermore, the fractional trend component is smoother than its $I(1)$ counterpart, as $\sigma_{\eta}$ in table \ref{ta:uni} shows. 
As \cite{KamMorWo2018} demonstrate, forcing the signal-to-noise ratio to be small can yield cycle estimates via correlated $I(1)$ UC models that are in line with economic theory. But as an inspection of their trend estimate shows, this comes with the cost of producing fractionally integrated long-run shocks that violate the white noise assumption. In contrast, fractionally integrated UC models directly estimate a small signal-to-noise ratio without restricting parameters to a certain interval and yield long-run shocks that are $I(0)$. Thus, a high signal-to-noise ratio in $I(1)$ UC models can indicate a violation of the $I(1)$ assumption for the long-run component. 
As explained in \cite{Web2011}, the strong negative correlation between $\eta_t$ and $\varepsilon_t$ is typically interpreted as causal impact from long-run shocks to the transitory component, where a positive trend shift yields a negative cyclical adjustment that vanishes over time due to the stationary nature of the transitory component. However, \cite{Web2011} also finds significant negative effects in the reverse direction.

\subsection*{Industrial production: plausible cycles in recessions}
\noindent
For log US industrial production we find $\hat{d}^{EW}=1.18$ and $\hat{d}^{GPH}=1.26$ which indicates a violation of the I(1) assumption of the unobserved components model. This is confirmed by the fractional trend-cycle model, for which the ML estimator yields $\hat{d}^{FT-FC}=1.66$. As figure \ref{fig:ll_gdp} shows, the likelihood is steep around $\hat{d}$. 

\begin{figure}[h!]
	\includegraphics[scale=0.82, trim = {0cm, 1.2cm, 0cm, 0cm}]{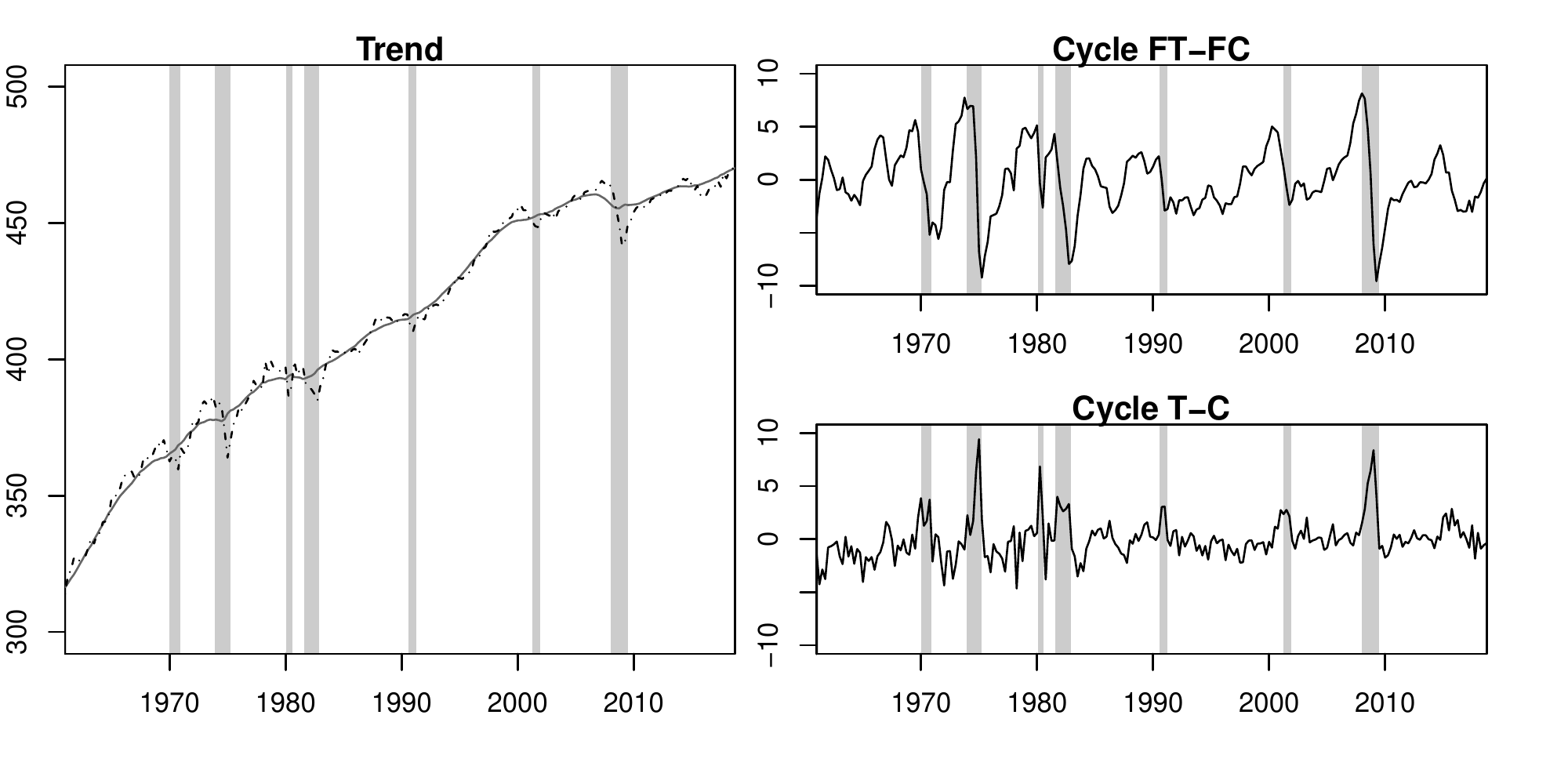}
	\caption[INDPRO: Trend Cycle]{Trend-cycle decompositions for log US real industrial production with correlated innovations. The left plot sketches the trend component estimate from the restricted model \eqref{eq:res} (T-C, with $d=1$) in black, dashed, together with the trend component from the unrestricted model \eqref{eq:unres} (FT-FC, with $d \neq 1$ allowed) in gray, solid. The plots on the right-hand side show the cyclical components for the unrestricted and the restricted model. Shaded areas correspond to NBER recession periods.}
	\label{fig:ip}
\end{figure}%

\noindent
Figure \ref{fig:ip} plots the unobserved components estimates from the T-C and the FT-FC specification for log US industrial production. Since $\hat{d}^{FT-FC}=1.66$ is considerably large, whereas $\hat{\sigma}_\eta^{{FT-FC}} = 0.14$ is relatively small compared to the benchmark $\hat{\sigma}_\eta^{{T-C}} = 8.09$, the fractional trend-cycle decomposition yields a smooth trend that only slightly drops during economic recessions, whereas the $I(1)$ counterpart is more erratic. The small ratio $\hat{\sigma}_\eta^{FT-FC}/\hat{\sigma}_\varepsilon^{FT-FC}$ may serve as an explanation for the differences between $\hat{d}^{FT-FC}$ and the nonparametric estimates $\hat{d}^{EW}$, $\hat{d}^{GPH}$, since a small signal-to-noise ratio can downward-bias the latter estimators \citep[cf.][]{SunPhi2004}.

\noindent
The cyclical component from the FT-FC specification is in line with the one obtained for log US real GDP, as it captures the dynamics from the business cycle well. It sharply drops during the NBER recession periods and recovers continuously in the aftermath, whereas the T-C cycle tends to increase during economic recessions, thereby contradicting economic theory. The results of \cite{Web2011}, who shows that a sufficiently long AR polynomial (in his case $p=10$ for monthly industrial production) can produce a more plausible cycle in an $I(1)$ correlated UC setup, are in line with the fractional cycle specification that can be interpreted as an autoregressive process of order $n-1$. 

\subsection*{Investment: strong cyclical variation}
\noindent
Turning to log US real gross private domestic investment, the nonparametric estimators yield $\hat{d}^{EW}=1.12$ and $\hat{d}^{GPH}=1.15$, whereas the maximum likelihood estimator for the fractional trend-cycle decomposition returns a slightly larger $\hat{d}^{FT-FC}=1.28$ as shown in table \ref{ta:uni}. The likelihood is relatively steep around $\hat{d}^{FT-FC}=1.28$, as figure \ref{fig:ll_gdp} indicates.
\begin{figure}[h!]
	\includegraphics[scale=0.82, trim = {0cm, 1.2cm, 0cm, 0cm}]{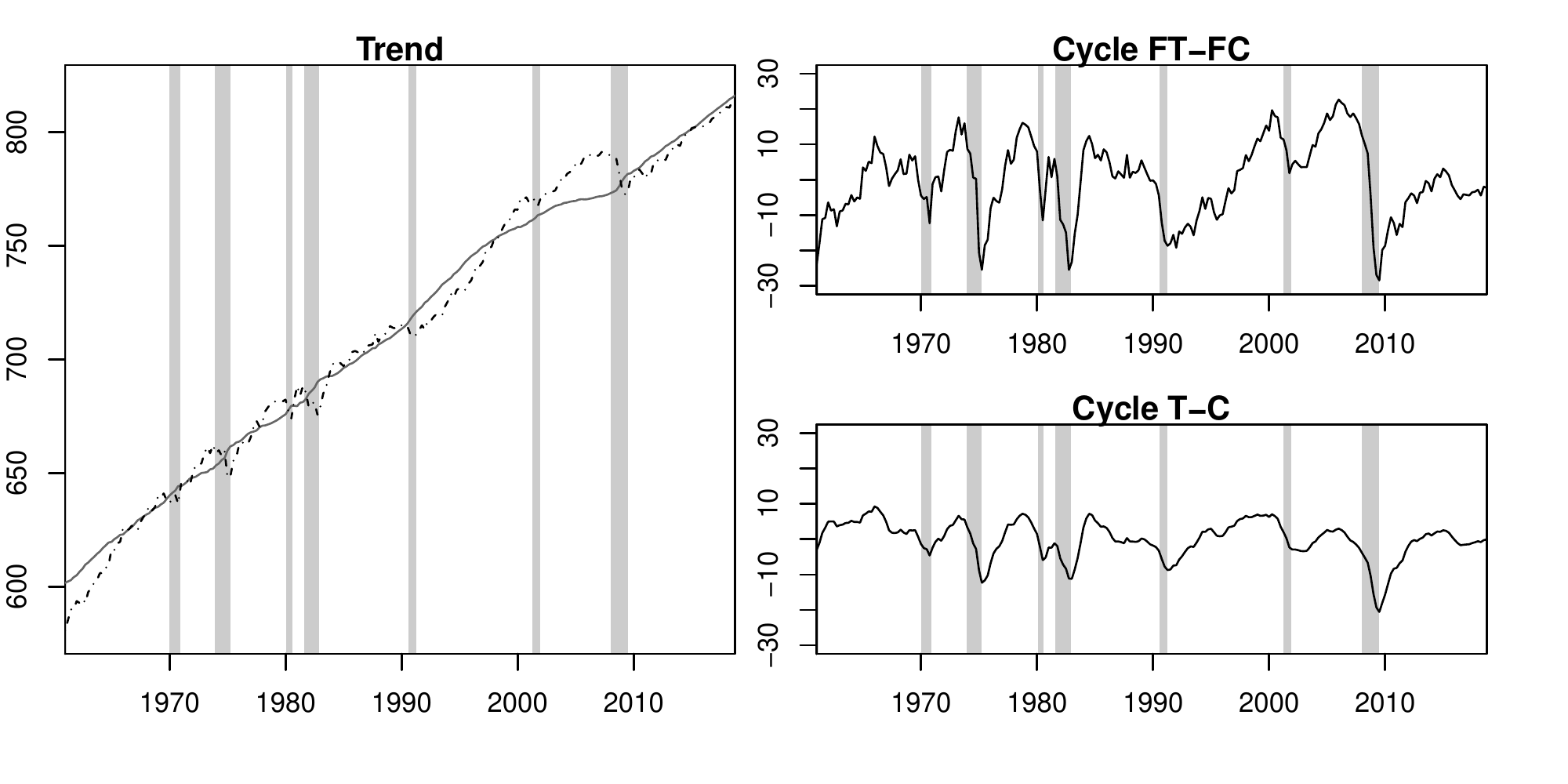}
	\caption[GPDIC1: Trend Cycle]{Trend-cycle decompositions for log US real gross private domestic investment with correlated innovations. The left plot sketches the trend component estimate from the restricted model \eqref{eq:res} (T-C, with $d=1$) in black, dashed, together with the trend component from the unrestricted model \eqref{eq:unres} (FT-FC, with $d \neq 1$ allowed) in gray, solid. The plots on the right-hand side show the cyclical components for the unrestricted and the restricted model. Shaded areas correspond to NBER recession periods.}
	\label{fig:invest}
\end{figure}%

\noindent
Figure \ref{fig:invest} shows that the FT-FC specification produces a smoother trend than the T-C benchmark and attributes a larger fraction of total variation to the cyclical component. More in line with economic theory, long-run investment from the FT-FC model is almost linear during economic upswings, whereas the T-C estimate peaks directly before the NBER recession periods. This is especially striking in the 2000s. There, the T-C model ascribes a permanent character to development before and in the great recession. The FT-FC model instead finds a strong cyclical upswing before the great recession, followed by a pronounced slump of the cycle. Regarding the debate on the nature and effects of the recession, this leads to clearly different conclusions.

\subsection*{Consumption: smooth trend}
\noindent
Finally, for log US real personal consumption we estimate $\hat{d}^{EW}=1.40$ and $\hat{d}^{GPH}=1.37$ via the nonparametric estimators. Similarly, the fractional trend-cycle model yields $\hat{d}^{FT-FC}=1.44$, as table \ref{ta:uni} shows.

\begin{figure}[h!]
	\includegraphics[scale=0.82, trim = {0cm, 1.2cm, 0cm, 0cm}]{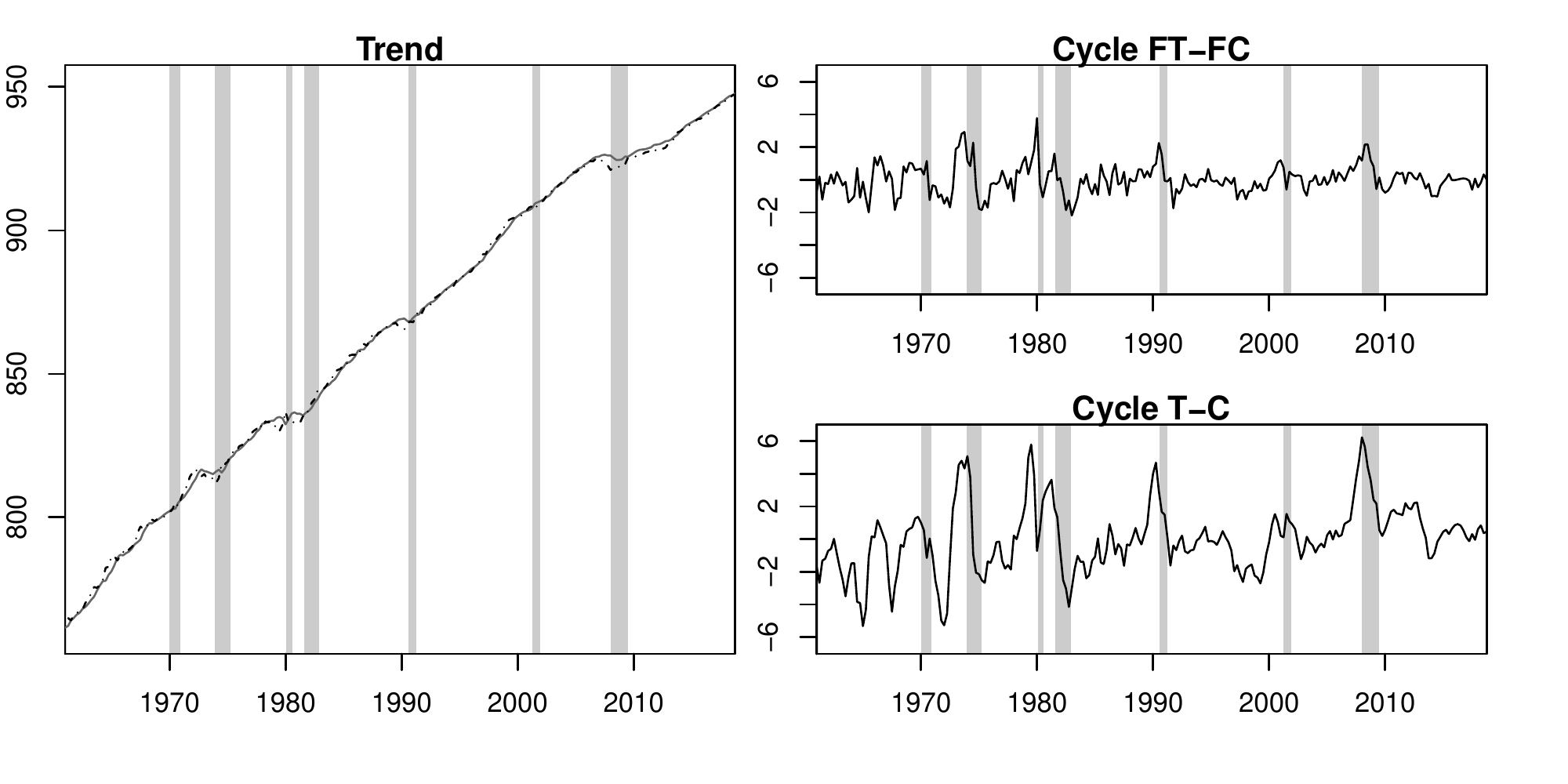}
	\caption[PCECC96: Trend Cycle]{Trend-cycle decompositions for log US real personal consumption expenditures with correlated innovations. The left plot sketches the trend component estimate from the restricted model \eqref{eq:res} (T-C, with $d=1$) in black, dashed, together with the trend component from the unrestricted model \eqref{eq:unres} (FT-FC, with $d \neq 1$ allowed) in gray, solid. The plots on the right-hand side show the cyclical components for the unrestricted and the restricted model. Shaded areas correspond to NBER recession periods.}
	\label{fig:pcecc96}
\end{figure}%

\noindent
Contrary to the results obtained for investment, the fractional decomposition attributes less variation to the cyclical component than the T-C benchmark. Hence, transitory consumption is estimated to be less volatile over the business cycle in the FT-FC framework. We find this more to be in line with economic theory than the results obtained from the T-C model, which indicate excessive overconsumption directly before a recession period.

\subsection*{Structural breaks and longer cycles}
Since \cite{PerWad2009} find that the stochastic long-run component of US GDP is well described by an I(0) process when a trend break in 1973:1 is introduced, we check the impact of the \cite{PerWad2009} break on our fractional trend-cycle decomposition. \cite{DieIno2001} argue that structural breaks and fractional trends can easily be confused. Hence, the robustness check clarifies whether the better performance of the fractional trend-cycle decomposition results from an ignored trend break. 

\noindent
Table \ref{ta:sb} reports the parameter estimates when a trend break in 1973:1 is allowed. As it shows, neither the integration order estimates $\hat{d}$, nor the autoregressive parameters and variance parameters differ substantially. The correlation between long- and short-run shocks is estimated to be slightly weaker when a trend break is introduced. The likelihood ratio (LR) test suggests that introducing a structural break in 1973:1 does not significantly improve the goodness of fit for GDP (p-value: $0.05$), industrial production (p-value: $0.12$), investment (p-value: $0.38$), and personal consumption (p-value: $0.09$).

\noindent
The trend-cycle decompositions in figures \ref{fig:gdp} -- \ref{fig:pcecc96} remain largely unaffected by the structural break, as figure \ref{fig:gdp:sb} in appendix \ref{app:gra} shows. 

\noindent
 As a second robustness check, we include further lags to the cyclical polynomial by setting $p=4$. In a non-fractional setting this implies that the cycle component contains lagged information from four quarters, which we consider as the maximum lag length of a cyclical component for quarterly data. By adding additional lags to the cyclical polynomial, we investigate if an increased flexibility of the cycle yields the same integration order estimates, or if the estimated fractional integration orders $\hat{d}$ are just an artifact from a too restrictive parametrization of $c_t$.
 Estimation results are given in table \ref{ta:rob} in appendix \ref{app:gra}. For industrial production, investment, and consumption additional lags have no significant impact. For GDP, slightly different autoregressive coefficients for the cycle are obtained, but they do not increase the overall fit of the model significantly, as a comparison of the likelihoods shows. Furthermore the estimated integration order is quite similar. The trend-cycle decompositions are sketched in figure \ref{fig:robust}. Since differences between the decompositions presented above and those contained in figure \ref{fig:robust} are negligible, we conclude that our results are robust to additional lags of the cyclical lag polynomials. 

\section{Conclusion} \label{Ch:4}
We generalized unobserved components models to the fractional domain by modelling the long-run component as a fractionally integrated series together with a cyclical component where the fractional lag operator enters the lag polynomial. We derived the reduced form representation, related the model to the decomposition of \cite{BevNel1981}, and showed that the model is uniquely identified independent of the lag length of the cyclical polynomial for $d \neq 1$. With the modified Kalman filter for the truncated, approximation-corrected state space representation of our fractional UC model we proposed a computationally feasible exact estimator for the latent components. 

\noindent
In an application to various macroeconomic series for the US, estimates for the cyclical component from the fractional trend-cycle model were often found to better capture the business cycle dynamics than those of a benchmark correlated unobserved components model with an $I(1)$ trend. E.g.\ for industrial production, the fractional trend-cycle model was shown to produce a cycle that is in line with economic theory. Furthermore, the fractional UC models estimated a smoother trend.
The reason for the better performance of fractionally integrated UC models compared to $I(1)$ UC models is the smaller signal-to-noise ratio, i.e.\ the ratio of long- and short-run shock variances. For $d>1$, as in our four applications, a violation of the $I(1)$ assumption in $I(1)$ UC models causes an upward-biased estimate of the long-run shock variance, which results in a high signal-to-noise ratio and, therefore, in a volatile trend estimate together with a noisy cycle. In contrast, allowing for fractional trends adequately captures the long-run dynamics of the trend and yields a consistent estimate of the long-run shock variance. In addition, the fractional lag operator $L_d$ attributes a higher variance to the short-run shocks to arrive at the same cyclical variance as in the $I(1)$ benchmark for $d>1$, thereby lowering the signal-to-noise ratio in the fractionally integrated UC model. Thus, the relatively small signal-to-noise ratio in the fractional model produces smooth trend estimates together with persistent cycles that reflect macroeconomic common sense.

\noindent
The fractional trend-cycle model offers a variety of opportunities for future research. The model may be generalized to the multivariate case, where fractional trends of different persistence with correlated innovations are allowed. A multivariate fractional trend-cycle model would then allow to estimate common fractional trends of cointegrated variables and test for polynomial cointegration. Furthermore, inferential methods that test for the number of common trends or the equality of integration orders could be established. As shown in \cite{DieIno2001}, fractionally integrated processes and structural breaks are related, since the former class of processes can produce level shifts and since structural breaks can be misinterpreted as $I(d)$ processes. Hence, combining both concepts, e.g.\ in a fractional UC model with regime switching, can be a fruitful challenge for future research. 

\noindent
To applied researchers, 
the model offers a flexible data-driven method to treat permanent and transitory components in macroeconomic and financial applications. It provides a solution for many issues of model specification that caused uncertainty and debates about realistic trend-cycle decompositions and estimation of recessions. Based on that, also the interaction of trends and cycles can be analyzed.

\clearpage
\appendix
\section{Graphs and tables} \label{app:gra}
\begin{table}[ht]
	\centering
	\begin{tabular}{lrrrrrrrr}
		& \multicolumn{2}{c}{GDP} &  \multicolumn{2}{c}{ind. production}  & \multicolumn{2}{c}{investment}  & \multicolumn{2}{c}{consumption}
		\\
		
		& T-C & FT-FC & T-C & FT-FC & T-C & FT-FC  & T-C & FT-FC\\
		\hline
		d 			&  					& 	1.32 		& 							& 1.66 				&  							& 1.28				&  					& 1.44 \\ 
		&  	 			& 	(0.12) 		& 							& (0.18)			&  							& (0.08) 			&  				& (0.07)  \\ 
		$\phi_1$		 & 1.29 		& 0.68 			& 0.51 					& 0.80					& 1.61 					& 0.90				& 1.19 		& 			0.45			\\ 
		& (0.21) 		& (0.29)		& (0.11) 				&(0.16)  				& (0.15) 				& (0.08) 			& (0.28) 	& (0.16)		\\ 
		$\phi_2$	       & -0.58 	& 					&0.05  					& 							& -0.67 				& 							& -0.39		 		& 0.13 \\ 
		& (0.18) 		&					&(0.04) 				& 				 			& (0.15) 			& 			 				& (0.23) 		&  (0.05)\\
		$\phi_3$	         &				&					&							& 							& 							&						& 					& 0.06\\
		&				&				&								& 						 &								&						& 					& (0.02)\\ 
		$\sigma_\eta^2$ & 1.45 & 0.36 & 8.09 & 0.14 & 7.01 & 0.51 & 1.82 & 0.33 \\ 
		$\sigma_{\eta \varepsilon}$ & -0.95 & -0.60 & -4.71 & -0.45 & 2.57 & -2.11 & -1.41 & -0.39 \\ 
		$\sigma_{\varepsilon}^2$ & 0.65 & 1.06 & 2.75 & 1.71 & 1.29 & 16.14 & 1.14 & 0.47 \\ 
		$\rho$ & -0.98 		& -0.97 		& -1					& -0.92 			& 		0.85 				& -0.74 			& -0.98 &-0.99\\ 
		$\log L$ & -261.01 & -260.53 	& -370.50 			& -370.03		 & 		-629.06				& -629.10 		& -208.12 & -199.31\\ 
		\hline
	\end{tabular}
	\caption[Univariate UC estimation]{Estimation results for the trend-cycle decomposition for log US real GDP, log US real industrial production, log US real gross private domestic investment, and log US real personal consumption expenditures. \textbf{T-C} distinguishes between an I(1) trend and an autoregressive cycle, and \textbf{FT-FC} between a fractionally integrated trend and an autoregressive cycle with fractional lag operator. $\rho$ denotes correlation between permanent and transitory shocks. $\log L$ is the log likelihood. Standard errors are in parentheses.}
	\label{ta:uni}
\end{table}

\begin{table}[t]
	\centering
	\begin{tabular}{lrlrlrlrl}
		& \multicolumn{2}{c}{GDP} &  \multicolumn{2}{c}{ind. production}  & \multicolumn{2}{c}{investment}  & \multicolumn{2}{c}{consumption} \\
		& Est. & Std.Err. & Est.& Std.Err. & Est. & Std.Err. & Est. & Std.Err. \\ 
		\hline
		d & 1.26 & 0.16 & 1.65 & 0.10 & 1.27 & 0.09 & 1.43 & 0.07 \\ 
		$\phi_1$ & 0.76 & 0.22 & 0.82 & 0.08 & 0.92 & 0.04 & 0.44 & 0.15 \\ 
		$\phi_2$ &  &  &  &  &  &  & 0.13 & 0.04 \\ 
		$\phi_3$ &  &  &  &  &  &  & 0.06 & 0.02 \\ 
		$\sigma_\eta^2$ & 0.41 & & 0.12&  & 0.43&  & 0.35&  \\ 
		$\sigma_{\eta \varepsilon}$ & -0.71  & & -0.37 & & -1.56  & & -0.41 & \\ 
		$\sigma_{\varepsilon}^2$ & 1.27 & &1.66& & 15.61& & 0.50 &\\ 
		$ \rho$		& -0.97 &			& -0.85&		& -0.60 &			& -0.99	&			\\
		log $L$   &	-258.55 &		& -368.81&		& -628.72 &		&-197.89 & \\
		\hline
	\end{tabular}
	\caption[Univariate UC estimation]{Robustness check: Estimation results for the trend-cycle decomposition for log US real GDP, log US real industrial production, log US real gross private domestic investment, and log US real personal consumption expenditures with a trend break in 1973:1. $\rho$ denotes correlation between permanent and transitory shocks. $\log L$ is the log likelihood.}
	\label{ta:sb}
\end{table}

\begin{table}[t]
	\begin{tabular}{lrlrlrlrl}
		& \multicolumn{2}{c}{GDP} &  \multicolumn{2}{c}{ind. production}  & \multicolumn{2}{c}{investment}  & \multicolumn{2}{c}{consumption}\\
		& Est. & Std.Err. & Est. & Std.Err. & Est, & Std.Err. & Est. & Std.Err. \\ 
		\hline
		d & 1.22 & 0.09 & 1.71 & 0.14 & 1.05 & 0.41 & 1.47 & 0.08 \\ 
		$\phi_1$ 	 & 0.64 & 0.33 & 0.76 & 0.12 & 1.05 & 0.23 & 0.40 & 0.17 \\ 
		$\phi_2$ 	 & 0.28 & 0.18 & -0.04 & 0.04 & -0.06 & 0.16 & 0.14 & 0.08 \\ 
		$\phi_3$ 	 & -0.32 & 0.15 & 0.03 & 0.01 & -0.01 & 0.08 & 0.07 & 0.03 \\ 
		$\phi_4$ 	 & -0.05 & 0.13 & 0.00 & 0.01 & -0.06 & 0.08 & 0.01 & 0.01 \\ 
		$\sigma_\eta^2$ & 0.66 & &0.17 & &2.32 & &0.32 &\\ 
		$\sigma_{\eta \varepsilon}$ & -0.22 & &-0.52 & &-6.01 & &-0.37 &\\ 
		$\sigma_\varepsilon^2$ & 0.18 & &1.61 && 22.31 & &0.41 &\\ 
		$ \rho$		& -0.64 &			& -0.99&		& -0.84 &			& -1	&			\\
		log $L$   &	-260.32 &		& -365.04&		& -628.55 &		&-198.88 & \\
		\hline
	\end{tabular}
	\caption[Univariate UC estimation]{Robustness check: Estimation results for the trend-cycle decomposition for log US real GDP, log US real industrial production, log US real gross private domestic investment, and log US real personal consumption expenditures with four autoregressive lags. $\rho$ denotes correlation between permanent and transitory shocks. $\log L$ is the log likelihood.}
	\label{ta:rob}
\end{table}

\begin{figure}[h]
	\includegraphics[scale=0.79, trim = {0cm, 1cm, 0cm, 2cm}]{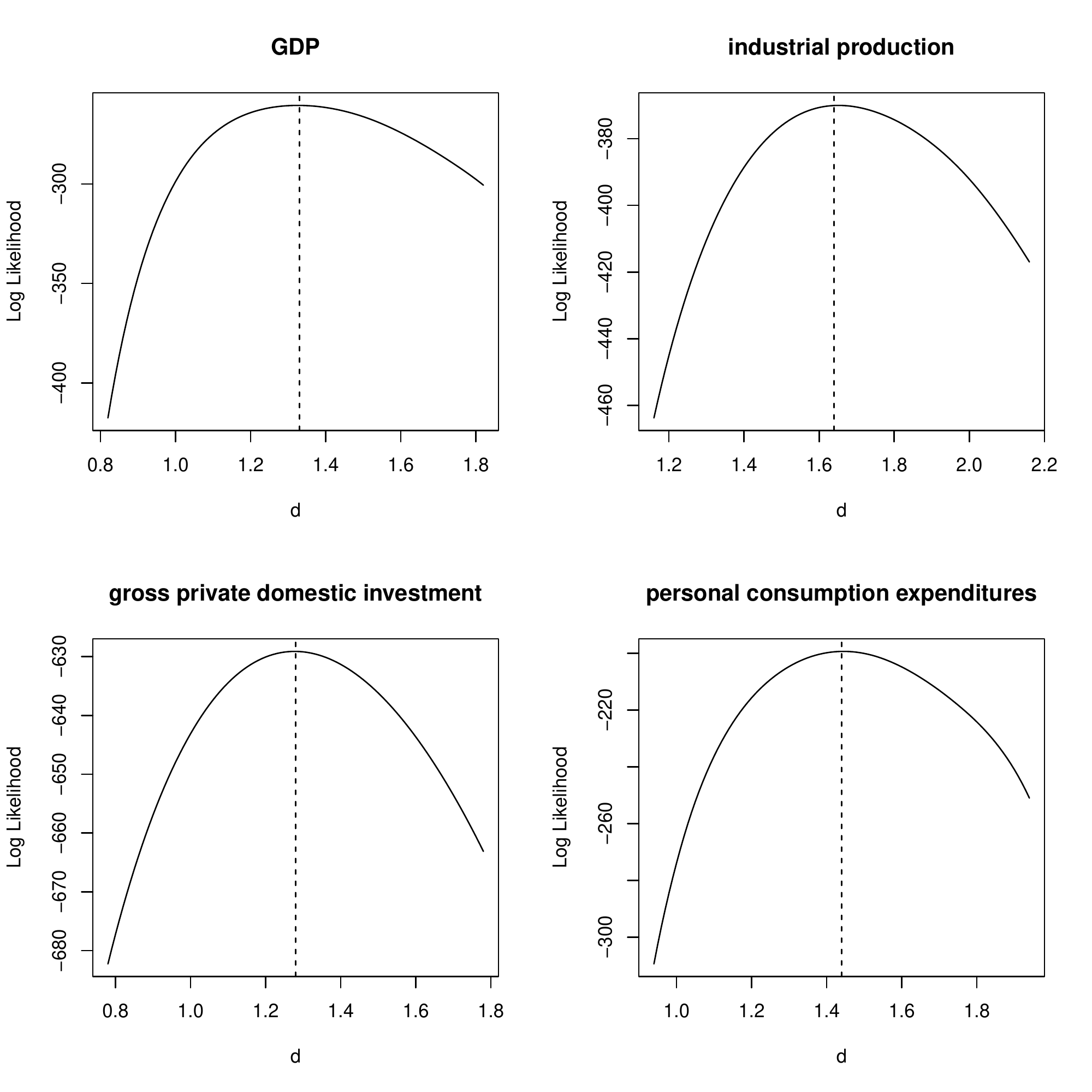}
	\caption[GDPC1: Log Likelihood]{Log Likelihood of the fractional trend-cycle decomposition of log US real GDP ($d \in [0.82, 1.82]$), log US industrial production ($d \in [1.16, 2.16]$), log US real gross private domestic investment ($d \in [0.78, 1.78]$), and log US real personal consumption expenditures ($d \in [0.94, 1.94]$). The remaining parameters in $\vtheta$ are fixed and given in table \ref{ta:uni}.}
	\label{fig:ll_gdp}
\end{figure}%

\begin{figure}[h]
	\includegraphics[scale=0.95, trim = {0cm, 1cm, 0cm, 0cm}]{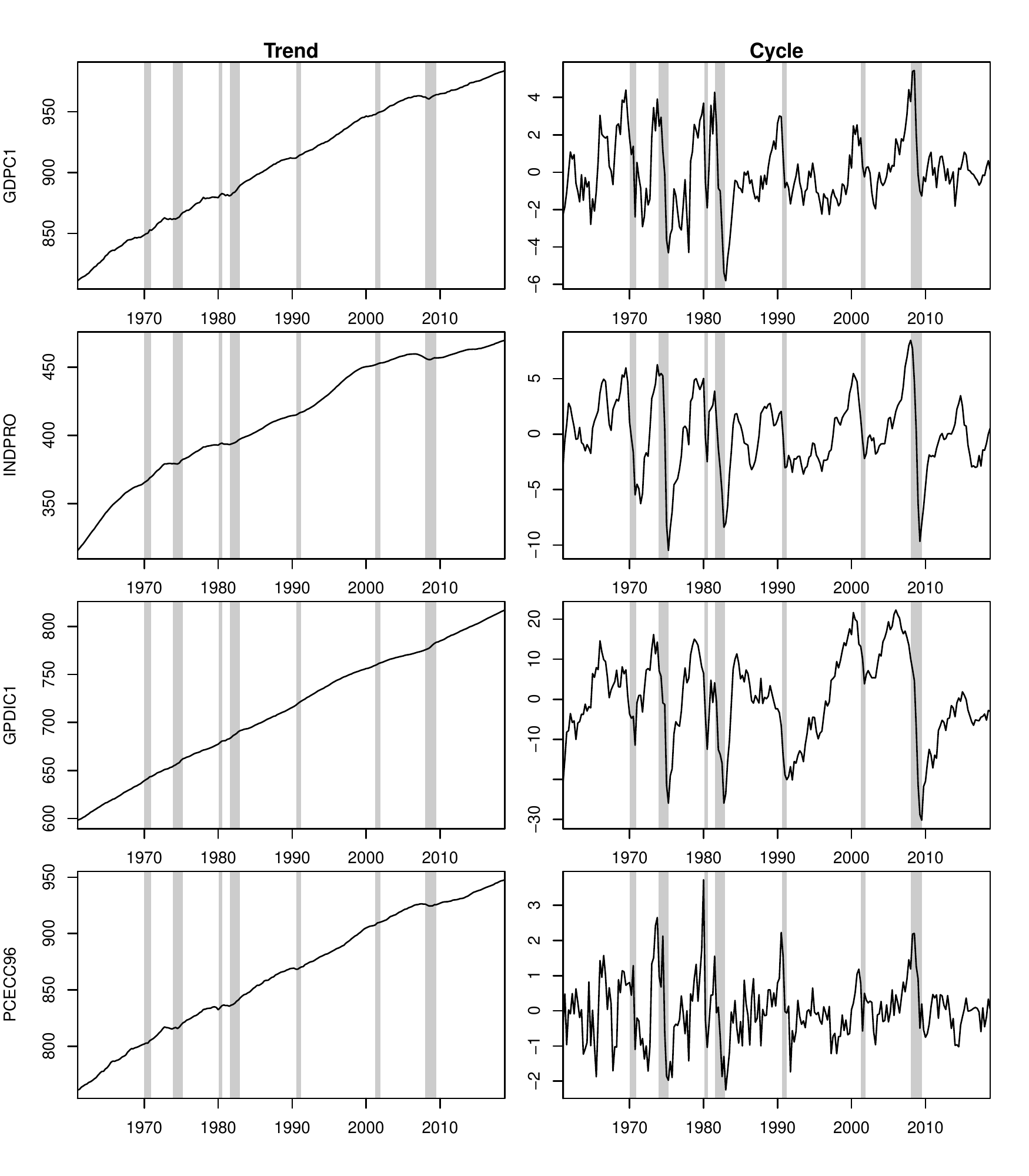}
	\caption[GDPC1: Trend Cycle]{Robustness check: Trend-cycle decompositions for log US real GDP (GDPC1), log US real industrial production (INDPRO), log US real gross private domestic investment (GPDIC1), and log US real personal consumption expenditures (PCECC96) with correlated innovations. The left plots sketch the trend component estimates from the unrestricted models \eqref{eq:unres} (FT-FC) with a trend break in 1973:1. The plots on the right-hand side show the cyclical components for the model with structural break. Shaded areas correspond to NBER recession periods.}
	\label{fig:gdp:sb}
\end{figure}%

\begin{figure}[h]
	\includegraphics[scale=0.95, trim = {0cm, 1cm, 0cm, 0cm}]{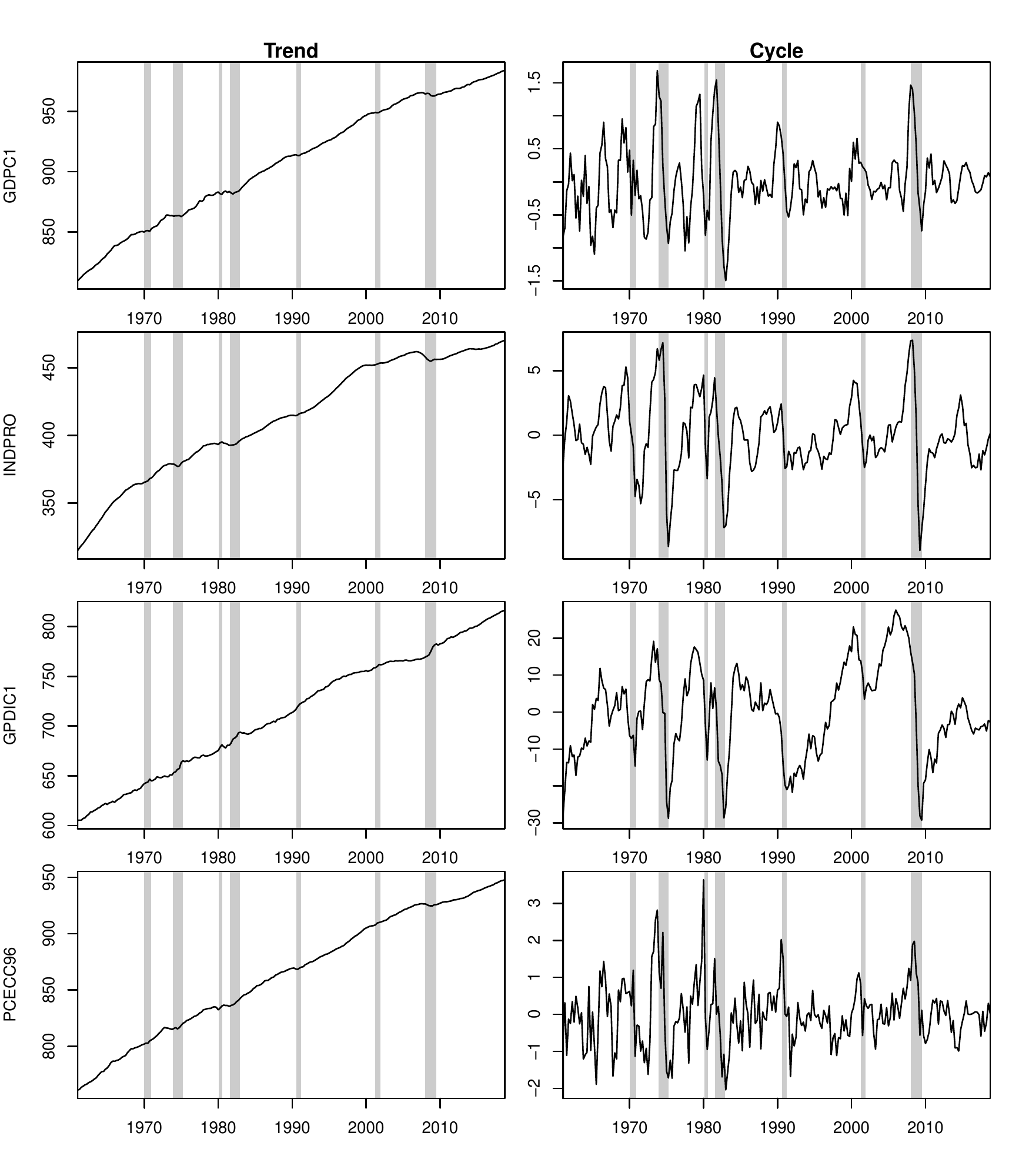}
	\caption[GDPC1: Trend Cycle]{Robustness check: Trend-cycle decompositions for log US real GDP (GDPC1), log US real industrial production (INDPRO), log US real gross private domestic investment (GPDIC1), and log US real personal consumption expenditures (PCECC96) with correlated innovations. The left plots sketch the trend component estimates from the unrestricted models \eqref{eq:unres} (FT-FC). The plots on the right-hand side show the cyclical components with four autoregressive lags. Shaded areas correspond to NBER recession periods.}
	\label{fig:robust}
\end{figure}%
\clearpage
\section{Univariate moving average representation of aggregated model}\label{App:MA_agg}

We consider the aggregation of two moving average (MA) processes in the lag operator $L_d$ with generic lag polynomials $h(L_d)$ and $ \tilde{h}(L_d)$ of order $q$ and $\tilde q$, respectively, 
\begin{equation} \label{eq:app_z_eta_eps}
z_t = h(L_d) \eta_t + \tilde{h}(L_d) \varepsilon_t ,
\end{equation}
with the white noise processes
\begin{equation}\notag
 	\begin{pmatrix}
		\eta_{t} \\ \varepsilon_{t} 
 	\end{pmatrix} \sim  \mathrm{i.i.d.}(0, \mQ), \qquad 
	 \mQ = \begin{bmatrix}
	 	\sigma_\eta^2 & \sigma_{\eta \varepsilon} \\
	 	\sigma_{\eta \varepsilon} & \sigma_\varepsilon^2
 \end{bmatrix}.
 \end{equation}
In what follows, set $p = \max (q, \tilde q)$ and let $h_i = 0$ for all $i > q$, $\tilde{h}_i = 0$ for all $i > \tilde{q}$. We first derive the MA representation in the standard lag operator $L=L_1$. Next we derive the MA representation in the fractional lag operator $L_d$ 
which is in general not of finite order.

\noindent
To rewrite  \eqref{eq:app_z_eta_eps} in the conventional lag operator $L$ define
\begin{align}\notag
L_d^k = (1-\Delta_+^d)^k = \left(\sum_{i= k}^\infty \varsigma_{k,i}(d)L^i \right)_+,
\end{align}
insert it into  \eqref{eq:app_z_eta_eps}, and rearrange terms
\begin{align} \nonumber
z_t & =\eta_t + \varepsilon_t +  \sum_{k=1}^p \left( h_k \sum_{i= k}^{t - 1} \varsigma_{k,i}(d) \eta_{t -i} + 
		\tilde{h}_k \sum_{i= k}^{t - 1} \varsigma_{k,i}(d) \varepsilon_{t-i} \right) \\
     &  =\eta_t + \varepsilon_t + \sum_{k=1}^p \sum_{i= k}^{t - 1} 	\varsigma_{k,i}(d) \left( h_k \eta_{t-i} + \tilde{h}_k \varepsilon_{t-i} \right). \nonumber \\
\intertext{Redefining the sum indexes we obtain}
   z_t  & = \eta_t + \varepsilon_t + \sum_{l=1}^{t-1} \eta_{t - l} \left(\sum_{k=1}^l \varsigma_{k,l}(d)  h_k \right)  +  \sum_{l=1}^{t-1}\varepsilon_{t - l}
     \left(\sum_{k=1}^l \varsigma_{k,l}(d)  \tilde{h}_k \right) \label{eq:app_m_tilde_m_detail} \\
     & = \sum_{l=0}^{t-1} g_l \,\eta_{t-l} + \sum_{l=0}^{t-1} \tilde{g}_l \, \varepsilon_{t-l},
     \label{eq:app_z_L_eta_eps}  
\end{align}
with $g_0=\tilde{g}_0=1$ and $g_l=\sum_{k=1}^l \varsigma_{k,l}(d)  h_k$ and $\tilde{g}_l=\sum_{k=1}^l \varsigma_{k,l}(d)  \tilde{h}_k$, $l=1,2,\ldots,t-1$. Note that both moving average processes are of order $n-1$ for a given sample size $n$. If \eqref{eq:app_z_L_eta_eps} can be aggregated, there exists a univariate moving average process of order less or equal to $n-1$
\begin{equation} \notag 
z_t = c(L) u_t, \quad u_t\sim i.i.d.(0,\sigma_u^2).
\end{equation}
To compute the coefficients $c_i$, note that $\Cov (z_t, c_l u_{t-l}) = \Cov ( z_t, g_l \eta_{t-l} + \tilde{g}_l \varepsilon_{t-l})$, which gives
\begin{equation} \label{eq:app_c_l}
c_l^2\sigma_u^2 = g_l^2 \sigma_{\eta}^2 + \tilde{g}_l^2 \sigma_{\varepsilon}^2 + 2g_l\tilde{g}_l \sigma_{\eta\varepsilon},\quad l=0,1,\ldots,t-1.
\end{equation}
To make the dependence of $c_l^2$ on the parameters of the fractional moving average polynomials explicit insert $g_l$ and $\tilde{g}_l$ into \eqref{eq:app_c_l}. This delivers for $l \geq 1$
\begin{align}
c_l^2 \sigma_u^2&= 
\left( \sum_{k=1}^l \varsigma_{k,l}(d)  h_k \right)^2 \sigma_{\eta}^2 + 
\left( \sum_{k=1}^l \varsigma_{k,l}(d)  \tilde{h}_k \right)^2 \sigma_{\varepsilon}^2 +
2\left( \sum_{k=1}^l \varsigma_{k,l}(d)  h_k \right)  
\left( \sum_{k=1}^l \varsigma_{k,l}(d)  \tilde{h}_k \right) \sigma_{\eta\varepsilon} \nonumber \\
& = \sum_{k=1}^l \sum_{i=1}^l \varsigma_{k,l}(d) \varsigma_{i,l}(d)
\left( h_k h_i \sigma_{\eta}^2 + \tilde{h}_k \tilde{h}_i \sigma_{\varepsilon}^2 + 2 \sigma_{\eta\varepsilon} h_k \tilde{h}_i
\right), \label{eq:app_c_l_in_bs}
\end{align}
with $c_0=1$, $\sigma_u^2 = \sigma_\eta^2 + \sigma_\varepsilon^2 + 2 \sigma_{\eta \varepsilon}$. Solving for $c_l$ yields the MA coefficients for $u_t$.

\noindent
Next we derive the univariate moving average representation in the fractional lag operator which is typically of infinite order
\begin{align} \label{eq:app_z_L_d_u}
z_t & = \psi_+(L_d) u_t.
\end{align}
If \eqref{eq:app_z_L_d_u} exists, then it can be rewritten  similarly to \eqref{eq:app_m_tilde_m_detail}  in the standard lag operator as
\begin{align}
z_t & = u_t +  \sum_{l=1}^{t-1} u_{t - l} \left(\sum_{k=1}^l \varsigma_{k,l}(d)  \psi_k \right).
\notag 
\end{align}
For such a representation to exist, there must exist parameters $\psi_i$, $i=1,\ldots, q_u$ such that 
\begin{align*}
c_l & = \sum_{k=1}^l \varsigma_{k,l}(d)  \psi_k, \quad l=1,2,\ldots,t-1,  
\end{align*}
while  \eqref{eq:app_c_l} holds. Solving for $\psi_l$ delivers
\begin{align} \label{eq:app_psi_l}
\psi_l & = \frac{c_l - \sum_{k=1}^{l-1} \varsigma_{k,l}(d)  \psi_k}{\varsigma_{l,l}(d)}.
\end{align}
Obviously, the order of the moving average polynomial in the fractional lag operator would only be of finite order $q_u$ if 
\begin{align}
c_l = \sum_{k=1}^{l-1} \varsigma_{k,l}(d)  \psi_k,\quad l > q_u.
\end{align}
In general this is not the case. In order to represent the $\psi_l$, $l=1,...,q_u$, in terms of the parameters $h_j$, $j=1,...,q$, and $\tilde{h}_k$, $k=1,...,\tilde{q}$, of the moving average polynomials in $L_d$, one inserts \eqref{eq:app_c_l_in_bs} into \eqref{eq:app_psi_l} and obtains
\begin{align}\label{eq:psi}
\psi_l & = 
\frac{ \sqrt{  \sum_{k=1}^l \sum_{i=1}^l \varsigma_{k,l}(d) \varsigma_{i,l}(d)
\left( h_k h_i \sigma_{\eta}^2 + \tilde{h}_k \tilde{h}_i \sigma_{\varepsilon}^2 + 2 \sigma_{\eta\varepsilon} h_k \tilde{h}_i \right) }  / \sigma_u - \sum_{k=1}^{l-1} \varsigma_{k,l}(d)  \psi_k}{\varsigma_{l,l}(d)}.
\end{align}
Since only $\psi_1,...,\psi_{l-1}$ enter \eqref{eq:psi}, $\psi_l$ can be calculated recursively, where the first coefficient is $\psi_1 = \sigma_u^{-1}\sqrt{h_1^2 \sigma_\eta^2 + \tilde{h}_1^2 \sigma_\varepsilon^2 + 2 h_1 \tilde{h}_1 \sigma_{\eta \varepsilon}}$ and $\sigma_u = \sqrt{\sigma_\eta^2 + \sigma_\varepsilon^2 + 2 \sigma_{\eta \varepsilon}}$.

\section{State space representation}\label{App:sta}
In this section we derive a state space representation of the univariate fractional trend plus cycle model in \eqref{eq:t}. Since for fixed sample size $n$ every fractionally integrated process of type II exhibits a finite-order autoregressive representation of length $n-1$, an exact state space form of the system \eqref{eq:t}, \eqref{eq:tau}, and \eqref{eq:c} exists, but is computationally infeasible for large $n$, as discussed at the beginning of section \ref{sec:ss}. As a solution, we derive an approximate version of the system \eqref{eq:t}, \eqref{eq:tau}, and \eqref{eq:c} and directly correct for the resulting approximation error. Define
\begin{align}
	\tilde{y}_t &= \tilde{\tau}_t + \tilde{c}_t \label{eq:trunc} \\
	\tilde{\tau}_t &= \mu_0  + \mu_1 t + \tilde{x}_t,  &&\tilde{x}_t = [a(L, d)^{-1}m(L, d)]_+ \eta_t = b_+(L, d) \eta_{t}, \label{eq:trunc_x} \\
	\tilde{\delta}_+(L, d, \phi)\tilde{c}_t &= \varepsilon_t, &&\tilde{c}_t = [\tilde{\delta}(L, d, \phi)^{-1}]_+\varepsilon_t  = \tilde{\omega}_+(L, d, \phi) \varepsilon_t, \label{eq:trunc_c}
\end{align}
where the approximation errors for \eqref{eq:trunc_x} and \eqref{eq:trunc_c} are given in \eqref{approximation:x} and \eqref{approximation:c}, $a(L, d)$ and $m(L, d)$ are the ARMA(v, w) polynomials that approximate the fractional difference operator in \eqref{eq:tau} and $\tilde{\delta}(L, d, \phi)$ truncates the fractional lag polynomial $\phi(L_d) = \sum_{i=0}^{p} \phi_i L_d^i = \sum_{i=0}^{\infty} \delta_i L^i$ in \eqref{eq:c} after lag $l$. Note that from \eqref{apx}, \eqref{apc} it follows that $\E_\vtheta (x_{t+1} - \tilde{x}_{t+1} | \mathcal{F}_t) = \epsilon_t^x$, $\E_\vtheta (c_{t+1} - \tilde{c}_{t+1} | \mathcal{F}_t) = \epsilon_t^c$, and consequently $\E_\vtheta(\tilde{y}_{t+1} | \mathcal{F}_t) = \E_\vtheta({y}_{t+1} | \mathcal{F}_t) -\epsilon_t^\tau - \epsilon_t^c = \E_\vtheta(\ddot{y}_{t+1} | \mathcal{F}_t) $ as defined in section \ref{sec:ss}. 

\noindent
The state equation for the stochastic long-run component $\tilde{x}_{t}$ is then given by
\begin{align*}
	\valpha_{t}^{x} = \bvec \tilde{x}_{t} \\ \tilde{x}_{t-1} \\ \vdots \\ \tilde{x}_{t-u+1} \\ \tilde{x}_{t-u} \evec= \bmat a_1  & 1 & 0 & \cdots & 0 \\
										a_2 & 0 & 1 & \cdots & 0 \\
										\vdots & \vdots & \vdots & \ddots & \vdots \\
										a_{u-1} & 0 & 0 & \cdots & 1\\
										a_u & 0 & 0 & \cdots & 0 \emat
										\bvec \tilde{x}_{t-1} \\ \tilde{x}_{t-2} \\ \vdots \\ \tilde{x}_{t-u} \\ \tilde{x}_{t-u-1} \evec + 
										\bvec 1 \\ m_1 \\ \vdots \\ m_{u-2} \\ m_{u-1} \evec \eta_{t}= \mT^x \valpha_{t-1}^x + \mR^x \eta_{t}, 
\end{align*}
where $u = \mathrm{max}(v, w+1)$. 

\noindent
The state equation for the cycle follows immediately 
\begin{align*}
	\valpha_{t}^c= \bvec \tilde{c}_{t}\\ \tilde{c}_{t-1} \\ \vdots \\ \tilde{c}_{t-l+1} \evec  = \bmat \tilde{\delta}_{1} & \cdots & \tilde{\delta}_{l-1} & \tilde{\delta}_{l} \\ 
									  1         & \cdots &  0 & 0\\
									  \vdots & \ddots & \vdots & \vdots\\
									  0 & \cdots & 1 & 0 \emat
			\bvec \tilde{c}_{t-1} \\ \tilde{c}_{t-2} \\ \vdots \\ \tilde{c}_{t-l}\evec  + 
			\bvec 1 \\ 0 \\ \vdots \\ 0 \evec \varepsilon_{t} = \mT^c \valpha_{t-1}^c + \mR^c \varepsilon_{t}. 
\end{align*}
Deterministic terms are incorporated as usual \citep[see, e.g.][ch. 3.2.1]{DurKoo2012} via $\valpha_{t}^{\mu}$.  \\
Finally, the observations equation is given by
\begin{align*}
	\tilde{y}_{t} = \bvec \mZ^\mu & \mZ^x & \mZ^c \evec \bvec \valpha_{t}^\mu \\ \valpha_{t}^x \\ \valpha_{t}^c \evec = \mZ \valpha_{t},
\end{align*}
where $\mZ^\mu = \bvec 1 & 0 \evec$, $\mZ^\tau = \bvec 1 & 0 &\cdots & 0 \evec$, and $\mZ^c= \bvec 1 & 0 & \cdots & 0\evec$. Since the Kalman filter estimates  $\E_\vtheta(\tilde{y}_{t+1} | \mathcal{F}_t) = \E_\vtheta(\ddot{y}_{t+1} | \mathcal{F}_t) $, and since the resulting prediction error is identical to the one of the exact representation as shown in \eqref{eq:v}, the maximum likelihood estimator for the unknown parameters $\vtheta$ based on the approximation-corrected truncated state space model \eqref{eq:trunc} - \eqref{eq:trunc_c} is identical to the one based on the exact representation \eqref{eq:t}, \eqref{eq:tau}, and \eqref{eq:c}. 

\clearpage
\begin{spacing}{1.2}
\bibliographystyle{dcu}
\bibliography{literatur.bib}
\end{spacing}
\end{document}